\documentclass[useAMS,usenatbib]{mn2e}

\pdfoutput=1

\usepackage{float}
\usepackage{enumerate}
\usepackage{multirow}
\usepackage{array}
\usepackage{comment}
\usepackage{ifthen}

\usepackage{amssymb,amsfonts,amsmath,graphicx,multirow}
\usepackage{graphicx}
\title[Asterodensity Profiling]{Characterizing Distant Worlds with Asterodensity Profiling}
\author[David M. Kipping]{David M. Kipping$^{1,2}$\thanks{E-mail:
dkipping@cfa.harvard.edu}\footnotemark[1]\\
$^{1}$Harvard-Smithsonian Center for Astrophysics, 60, Garden Street, Cambridge, MA 02138 \\
$^{2}$Carl Sagan Fellow}
\begin{document}

\date{Accepted 2014 February 12. Received 2014 January 10; in original form 2013 November 4}

\pagerange{\pageref{firstpage}--\pageref{lastpage}} \pubyear{2013}

\maketitle

\newcommand{\luna}{{\tt LUNA}}
\newcommand{\multi}{{\sc MultiNest}}
\newcommand{\macula}{{\tt macula.f90}}
\newcommand{\obs}{\mathrm{obs}}
\newcommand{\tru}{\mathrm{true}}
\newcommand{\blend}{\mathrm{blend}}

\label{firstpage}

\begin{abstract}

Eclipsing systems, such as transiting exoplanets, allow one to measure the mean 
stellar density of the host star under various idealized assumptions. 
Asterodensity Profiling (AP) compares this density to an independently 
determined value in order to check the validity of the assumptions and 
ultimately derive useful parameters. Several physical effects can cause said 
assumptions to become invalid, with the most well-known example being the 
so-called photo-eccentric effect. In this work, we provide analytic expressions 
for five other effects which induce AP deviations: the photo-blend, -spot, 
-timing, -duration and -mass effects. We find that these effects can easily 
reproduce large AP deviations and so we caution that extracting the eccentricity 
distribution is only viable with careful consideration of the prior 
distributions for these other effects. We also re-investigate the 
photo-eccentric effect and derive a single-domain minimum eccentricity 
expression and the parameter range for which analytic formulae are valid. This 
latter result shows that the assumptions underlying the analytic model for the 
photo-eccentric effect break down for close-in, highly-eccentric planets, 
meaning that extreme care must be taken in this regime. Finally, we demonstrate 
that contaminated light fraction can be solved for, indicating that AP could be 
a potent tool for planet validation.

\end{abstract}


%%%%%%%%%%%%%%%%%%%%%%%%%%%%%%%%%%%%%%%%%%%%%%%%%%%%%%%%%%%%%%%%%%%%%%%%%%%%%%%%
%%%%%%%%%%%%%%%%%%%%%%%%%%%% SECTION 1: INTRODUCTION %%%%%%%%%%%%%%%%%%%%%%%%%%%
%%%%%%%%%%%%%%%%%%%%%%%%%%%%%%%%%%%%%%%%%%%%%%%%%%%%%%%%%%%%%%%%%%%%%%%%%%%%%%%%

\section{Introduction}
\label{sec:intro}

Asterodensity profiling (AP) is a relatively new concept in the study of
astronomical eclipses, such as transiting planets and eclipsing binaries, with 
the potential to constrain various properties of an eclipsing system using 
photometric data alone. AP exploits a well-known trick in the field of 
photometric eclipses that if an object transits across the face of a star 
multiple times, then one can measure the mean density of the host star, 
$\rho_{\star}$, using Kepler's Third Law alone, under various idealized 
assumptions. This was first demonstrated in the pioneering work of 
\citet{seager:2003} and the most common application of this trick in the study
of exoplanets has been to use the $\rho_{\star}$ measurement as a luminosity
indicator for stellar evolution models, in order to obtain physical dimensions 
for the host star \citep{sozzetti:2007}.

AP goes further than this though, by comparing the transit light curve derived
stellar density, $\rho_{\star,\obs}$, to some independent measure of the same 
term, $\rho_{\star,\tru}$, in order to test the validity of the idealized
assumptions and ultimately extract information on the state the eclipsing 
system. If all of the idealized assumptions made in the definition of 
$\rho_{\star,\obs}$ are correct \citep{seager:2003}, then naturally one expects
$(\rho_{\star,\obs}/\rho_{\star,\tru})=1$ (to within the measurement 
uncertainties). Any deviation from unity implies that
one or more of the idealized assumptions are invalid and the magnitude and 
direction of this deviation provide insights into the physical origin of the
discrepancy. These idealized assumptions include (but are not limited to) an 
opaque planet, a spherical planet, a spherical star, non-variable transit shape, 
Keplerian circular orbit and negligible blending from unresolved luminous 
objects.

The first usage of the term ``asterodensity profiling'' was by \citet{map:2012},
who focussed on Multi-body Asterodensity Profiling (MAP) to constrain mutual
orbital eccentricities. By focussing on systems with multiple transiting 
planets, several measurements of $\rho_{\star,\obs}$ are obtained, allowing one 
to seek relative discrepancies in $\rho_{\star,\obs}$, rather than the absolute 
discrepancy determined when $\rho_{\star,\tru}$ is known. MAP is particularly 
powerful since it makes no assumption about the true stellar density. 

Although \citet{map:2012} briefly speculated that Single-body Asterodensity 
Profiling (SAP) would be plausible if a very tight constraint on 
$\rho_{\star,\tru}$ was available, such as that from asteroseismology, 
\citet{dawson:2012a} proposed that even a loose prior on $\rho_{\star,\tru}$
would be sufficient to identify highly eccentric planets. Referring to the
effect as the ``photo-eccentric effect'', the authors demonstrated the technique
on the known eccentric planet HD~17156b obtaining $e=0.71_{-0.09}^{+0.16}$
in good agreement with the radial velocity determination of $e=0.67\pm0.08$.
In later work, the same authors showed that the \emph{Kepler} planetary 
candidate KOI-1474.01 has an eccentricity of $e=0.81_{-0.07}^{+0.10}$, if the 
candidate is genuine \citep{dawson:2012b}. In the case of ostensibly 
near-circular orbits, SAP provides less constraining determinations; for example 
\citet{kepler22:2013} recently used SAP on Kepler-22b to determine 
$e=0.13_{-0.13}^{+0.36}$ (we propose an explanation for this in 
\S\ref{sub:photoeccentric}).

We also note that variants of AP have been explored in the exoplanet literature,
although they are not referred to as AP explicitly. Since $\rho_{\star,\obs}$ is 
a function of the observed transit durations (as shown later in 
\S\ref{sec:principles}), several previous works have re-phrased the problem by 
looking for anomalous transit durations (e.g. \citealt{moorhead:2011,
kane:2012}). One particularly powerful advantage of explicit AP is that 
$\rho_{\star,\tru}$ is a direct observable from the ``gold standard'' inference 
from asteroseismology, using the frequency spacing of pulsations modes 
\citep{ulrich:1986}. In contrast, the ``true'' transit duration or maximum 
transit duration can, in general, only be inferred by invoking stellar evolution 
models since one needs to estimate the stellar radius, $R_{\star}$ 
\citep{moorhead:2011}.

AP, and variants thereof, have so far been predominantly employed for 
constraining orbital eccentricities in both individual systems (e.g. \citealt{dawson:2012a,dawson:2012b,kepler22:2013}) and with regard to the entire 
eccentricity distribution (e.g. \citealt{moorhead:2011,kane:2012,dawson:2013}). 
The former goal has a particularly important place with regard to assessing 
habitability of planetary candidates since eccentricity can have severe effects 
\citep{dressing:2010}. The latter is mostly concerned with testing 
planet-formation models \citep{ford:2008,juric:2008,socrates:2012,dong:2013}.

The importance of measuring eccentricities is therefore apparent; thus 
explaining the recent focus of applying AP for constraining eccentricities
via the photo-eccentric effect. However, relatively little work exists in
the literature exploring the other physical effects which can lead to
$(\rho_{\star,\obs}/\rho_{\star,\tru})\neq1$. This absence of investigation is
problematic since a circular orbit is not the only idealized assumption in the
definition of $\rho_{\star,\obs}$ which may be in error, and thus responsible 
for an observation that $(\rho_{\star,\obs}/\rho_{\star,\tru})\neq1$. 
Critically, negating these other effects may lead to systematic errors in 
derived eccentricities or even completely erroneous conclusions about the state 
of a system. Additionally, the analytic expressions for AP are only approximate
forms \citep{map:2012}, and yet the explicit valid range for their applicability 
remains unknown. The purpose of this work is to provide analytic expressions for 
several plausible alternative mechanisms by which AP can produce 
$(\rho_{\star,\obs}/\rho_{\star,\tru})\neq1$ and define the exact
parameter range for which these expressions may be reasonably employed without a
significant loss of accuracy. We therefore aim to provide a foundational 
theoretical framework for this burgeoning field of study.

%%%%%%%%%%%%%%%%%%%%%%%%%%%%%%%%%%%%%%%%%%%%%%%%%%%%%%%%%%%%%%%%%%%%%%%%%%%%%%%%
%%%%%%%%%%%%%%%%%%%%%%%%%%%%% SECTION 2: PRINCIPLES %%%%%%%%%%%%%%%%%%%%%%%%%%%%
%%%%%%%%%%%%%%%%%%%%%%%%%%%%%%%%%%%%%%%%%%%%%%%%%%%%%%%%%%%%%%%%%%%%%%%%%%%%%%%%

\section{Principles of Asterodensity Profiling}
\label{sec:principles}

\subsection{Determining $\rho_{\star,\obs}$}
\label{sub:rhoobs}

It is not the purpose of this work to provide a detailed introductory review
of basic transit theory. Despite this, we here provide a brief synposis of how
the mean stellar density is derived in the context of the AP technique. Those
interested in a more detailed pedagogical discussion are directed to 
\citet{winn:2010}.

Throughout this work, including all appendices, we make the fundamental 
assumption that any observed transits satisfy the criteria $0<b<(1-p)$ where $b$ 
is the impact parameter of the transiting object of $p$ is the ratio-of-radii
between the transiting object and the host star. In the absence of 
limb-darkening, such a transit would be described as exhibiting a flat-bottom.
Since $b>0$ at times, then our fundamental assumption also enforces the condition
that $p<1$ at all times. This means our work does not include total eclipses 
caused by planets orbiting white dwarf stars for example \citep{agol:2011}, 
yet for which there are no observed examples to date. Employing this fundamental 
assumption, a transit provides four basic observational parameters:

\begin{itemize}
\item[{\tiny$\blacksquare$}] $\delta_{\obs}$: the observed transit depth
\item[{\tiny$\blacksquare$}] $\tau_{\obs}$: the observed time of transit minimum
\item[{\tiny$\blacksquare$}] $T_{14,\obs}$: the observed first-to-fourth contact 
transit duration
\item[{\tiny$\blacksquare$}] $T_{23,\obs}$: the observed second-to-third contact 
transit duration
\end{itemize}

The transit depth scales with the size of the transiting object and thus
$p_{\obs}$ is easily recovered. Multiple epochs provide several $\tau_{\obs}$ 
measurements which can be used to infer the orbital period of the transiting 
object, $P$. The other two observables, $T_{14,\obs}$ and $T_{23,\obs}$, may be 
used to determine the observed impact parameter, $b_{\obs}$, and the observed
scaled semi-major axis of the orbit, $(a/R_{\star})_{\obs}$, as demonstrated by 
\citet{seager:2003}. Under the assumption of a spherical, opaque, dark planet on 
a Keplerian circular orbit transiting a spherical, unblended host star, 
\citet{seager:2003} showed that the transit durations would be given by

\begin{align}
T_{ _{23}^{14} } &= \frac{P}{\pi} \sin^{-1} \Bigg[ \sqrt{ \frac{(1\pm p)^2 - b^2}{(a/R_{\star})^2-b^2} } \Bigg].
\label{eqn:durations}
\end{align}

These expressions may be solved simultaneously for $b$ and $(a/R_{\star})$. We 
refer to these expressions as the \emph{observed} impact parameter and 
\emph{observed} scaled semi-major axis since both terms are only valid under the 
various assumptions made thus far.

\begin{align}
b_{\obs}^2 &\equiv \frac{ (1-p_{\obs})^2 - \frac{\sin^2(T_{23,\obs}\pi/P)}{\sin^2(T_{14,\obs}\pi/P)} (1+p_{\obs})^2 }{ 1 - \frac{\sin^2(T_{23,\obs}\pi/P)}{\sin^2(T_{14,\obs}\pi/P)} }
\label{eqn:bseager}
\end{align}

\begin{align}
(a/R_{\star})_{\obs}^2 &\equiv \frac{ (1+p_{\obs})^2 - b_{\obs}^2 [1-\sin^2(T_{23,\obs}\pi/P)] }{ \sin^2(T_{23,\obs}\pi/P) }.
\label{eqn:aRseager}
\end{align}

It is now trivial to show that $\rho_{\star,\obs}$ is found using Kepler's Third 
Law:

\begin{align}
\rho_{\star,\obs} &\equiv \frac{ 3\pi (a/R_{\star})_{\obs}^3}{G P^2},
\label{eqn:rhoseager}
\end{align}

where $G$ is the Gravitational constant.
Equation~\ref{eqn:rhoseager} also assumes $M_P\ll M_{\star}$ in addition to the 
previous assumptions and we use the equivalent symbol since the above represents 
a definition which we will use throughout this work. Note, that we refer to this
density with the subscript ``obs'' for observed, whereas previous works have 
used the subscript ``circ'' for circular (e.g. \citealt{dawson:2012a,map:2012}). 
The reason for this change is that, as demonstrated throughout this paper, 
numerous other idealized assumptions are made to derive 
Equation~\ref{eqn:rhoseager} in addition to a circular orbit and it is somewhat
misleading to label the term with ``circ'' since it implies that this is the 
only relevant assumption.

It is important to stress that limb darkening parameters do not feature in the
calculation of $\rho_{\star,\obs}$. In other words, $\rho_{\star,\obs}$ is
not functionally dependent upon the limb darkening coefficients (LDCs) or 
profile; e.g. $\rho_{\star,\obs} \neq f(u_1,u_2)$ in the case of quadratic limb 
darkening. This can be understood on the basis that the LDCs do not affect the 
instant at which the planet's projected disc contacts the star's projected disc
i.e. the contact points, since this is purely dynamical. Therefore, the transit 
durations, $T_{14,\obs}$ and $T_{23,\obs}$, are not affected by the LDCs in 
anyway. Since $\rho_{\star,\obs}$ depends solely upon $p_{\obs}$, $P$, 
$T_{14,\obs}$ \& $T_{23,\obs}$, then $\rho_{\star,\obs}$ must also be 
independent of the LDCs. In practice, one could arrive at the wrong 
$\rho_{\star,\obs}$ by fixing the LDCs to some values which do not represent 
the truth. This would lead to a biased estimate of $p_{\obs}$, $T_{14,\obs}$ \& 
$T_{23,\obs}$, and consequently a biased estimate of $\rho_{\star,\obs}$. We 
therefore advocate careful selection of the priors in the LDCs and specifically
suggest employing the non-informative prior basis set proposed in 
\citet{LDfitting:2013}, which will propagate the uncertainty of the LDCs into
the derivation of $\rho_{\star,\obs}$. Essentially, this means that the derived 
$\rho_{\star,\obs}$ value loses precision but gains accuracy - a satisfactory
compromise in most cases. Having established that it makes no difference to any 
of the derivations in this work whether we include/exclude limb darkening, many 
of the figures in this paper will negate it for the sake of clarity but once 
again we stress that it does not affect the validity of the derived expressions.

Finally, we note that the reason why we earlier stated that we will assume 
$b<(1-p)$ at all times is evident from the above expressions, since $T_{23}$
is undefined otherwise and thus it is not possible to calculate 
$\rho_{\star,\obs}$. Therefore, using the approach of \citet{seager:2003}, one 
can only measure the light curve derived stellar density of a star if $b<(1-p)$ 
and thus AP is only possible in such a regime.

\subsection{Observations versus Truth}
\label{sub:obsVStru}

Ideally, the observed transit depth and durations are equivalent to the true
values. In such a case, one should expect (to within the measurement 
uncertainties) that

\begin{align}
\lim_{\mathrm{idealized\,\,assumptions\,\,valid}} \Big(\frac{\rho_{\star,\obs}}{\rho_{\star,\tru}}\Big) &= 1.\nonumber
\end{align}

However, as is shown in this work, there are many realistic conditions which
do not satisfy the ideal transit assumptions made in \citet{seager:2003}.
Rather than seeing this as nuisance though, the principle of AP is to exploit
the $(\rho_{\star,\obs}/\rho_{\star,\tru})$ ratio to not only test the validity
of the idealized assumptions but to actually infer properties of an eclipsing 
system by analysis of the magnitude and direction of any discrepancies (or lack 
there-of). 

As mentioned earlier, either an independent measure of $\rho_{\star,\tru}$ is 
required to perform the SAP variant or relative differences between multiple 
transiting object can be used to perform MAP.

\subsection{Methodology for Analytic Derivations of AP Effects}
\label{sub:method}

There are many different physical scenarios which can cause a significant
AP discrepancy (which we define as when $(\rho_{\star,\obs}/\rho_{\star,\tru})
\neq1$ at high significance). In this work, we attempt to derive analytic 
expressions for several important effects to aid observers interpreting such 
measurements. In general, an unaccounted for effect (dubbed a 
``photo-\emph{name} effect'' throughout this work) will cause a systematic and 
constant deviation in either the depth or the duration such that 
$p_{\obs}\neq p_{\tru}$, $T_{14,\obs}\neq T_{14,\tru}$ and/or $T_{23,\obs}\neq 
T_{23,\tru}$. Unaccounted-for periodic transit timing/duration/depth variations 
(TTV/TDV/T$\delta$V) induced by perturbing gravitational influences or 
starspots can be interpreted as a systematic, constant deviation in the 
composite transit light curve's durations and/or depth too, as shown later in 
\S\ref{sub:photospot}, \S\ref{sub:phototiming} \& \S\ref{sub:photoduration}. 
Therefore, in general, one may derive $\rho_{\star,\obs}$ by considering it to 
be functionally dependent via:

\begin{align}
\rho_{\star,\obs}[p_{\obs}(p_{\tru},\mathbf{X}),T_{14,\obs}(T_{14,\tru},\mathbf{X}),T_{23,\obs}(T_{23,\tru},\mathbf{X})],\nonumber
\end{align}

where $\mathbf{X}$ is a vector of arbitrary length representing the parameters
which describe the unaccounted-for physical effect(s). In practice, one
computes the expressions for $p_{\obs}$, $T_{14,\obs}$ and $T_{23,\obs}$
and then uses Equations~\ref{eqn:bseager}, \ref{eqn:aRseager} \&
\ref{eqn:rhoseager} to analytically express 
$\rho_{\star,\obs}(p_{\tru},b_{\tru},\rho_{\star,\tru})$. In practice, the
the derived expression is often extremely cumbersome and impractical and thus
the major challenge of such work is a) finding a simplified, useful 
approximate expression by invoking various assumptions b) determining the exact 
conditions for which the associated assumptions are valid. These two goals
and the described basic methodology guide the work which follows throughout this 
paper. In general, we do not provide detailed derivations in the main text
for the sake of brevity, but all relevant derivations are included in detail
in the appendices.

%%%%%%%%%%%%%%%%%%%%%%%%%%%%%%%%%%%%%%%%%%%%%%%%%%%%%%%%%%%%%%%%%%%%%%%%%%%%%%%%
%%%%%%%%%%%%%%%%%%%%%%%%%%%%%% SECTION 3: PHOTO-ALL %%%%%%%%%%%%%%%%%%%%%%%%%%%%
%%%%%%%%%%%%%%%%%%%%%%%%%%%%%%%%%%%%%%%%%%%%%%%%%%%%%%%%%%%%%%%%%%%%%%%%%%%%%%%%

\section{Asterodensity Profiling Effects}
\label{sec:photoall}

\subsection{The Photo-mass Effect}
\label{sub:photomass}

We begin our exploration of various AP effects by considering that the idealized
assumption $M_{\mathrm{transiter}}\ll M_{\star}$ is invalid (the masses of 
transiting object and star respectively). We note that \citet{dawson:2012a}
briefly commented on this possibility previously (see \S4.3). As with the 
subsequent sections, we will assume that all of the other idealized assumptions 
remain valid in order to derive the consequences of the ``photo-mass'' effect in 
isolation. In general, a confirmed exoplanet will safely satisfy this criteria 
but planetary candidates cannot so easily be treated, since the observations 
could be of an eclipsing binary or a white/brown-dwarf with a high mass ratio. 
Including the $M_{\mathrm{transiter}}$ term in the derivation of the stellar 
density returns the result

\begin{align}
\rho_{\star,\obs} &= \rho_{\star,\tru} + p^3 \rho_{\mathrm{transiter}},
\end{align}

where $\rho_{\mathrm{transiter}}$ is the mean density of the transiting object
(usually this is a transiting planet but the expressions are valid for
eclipsing binaries too). This result implies that

\begin{align}
\Bigg(\frac{\rho_{\star,\obs}}{\rho_{\star,\tru}}\Bigg) &= 1 + p^3 \frac{\rho_{\mathrm{transiter}}}{\rho_{\star,\tru}},
\end{align}

which we may re-express as

\begin{align}
%\mathbf{The\,\,}&\mathbf{Analytic\,\,Photomass\,\,Effect} \nonumber \\
&\Bigg(\frac{\rho_{\star,\obs}}{\rho_{\star,\tru}}\Bigg)^{\mathrm{PM}} = 1 + \frac{M_{\mathrm{transiter}}}{M_{\star}},
\label{eqn:photomass}
\end{align}

where we use the subscript ``PM'' as an acronym for the photo-mass effect.
Negating the planetary mass therefore causes us to overestimate 
$(\rho_{\star,\obs}/\rho_{\star,\tru})$. For a confirmed/validated exoplanet, 
this effect will be $\lesssim 1$\% and thus is usually only minor. For eclipsing
binaries masquerading as planetary candidates through blending, this effect
will become order unity.

\subsection{The Photo-blend Effect}
\label{sub:photoblend}

One of the most critical assumptions in the derivation of $\rho_{\star,\obs}$
is that the brightness variations observed are due to the host star alone,
which means that the star is unblended. Blend sources come in many varieties
involving triple and binary stellar configurations \citep{torres:2011,
hartman:2011} as well as even self-blending due to a hot compact object such as 
a white-dwarf or even a hot-Jupiter \citep{nightside:2010}. Blend sources are 
the astrophysical bottleneck in confirming/validating the thousands of planetary 
candidates found by the \emph{Kepler Mission} \citep{morton:2011,fressin:2011}.

We define the blend factor, $\mathcal{B}$, in this work as the ratio of the 
total flux to that of the target's flux, via

\begin{align}
\mathcal{B} &\equiv \frac{ F_{\star} + F_{\blend} }{ F_{\star} },
\label{eqn:Beqn}
\end{align}

where $F_{\star}$ is the flux received from the target and $F_{\blend}$ is the 
sum of all extra contaminating components. In Appendix~\ref{app:photoblend},
we show that if we assume $(a/R_{\star})^2 \gg (1+p)^2$, the effect of a blend 
may be expressed as (see Equation~\ref{eqn:rhoobs2}):

\begin{align}
%\mathbf{The\,\,Analytic}&\mathbf{\,\,Photoblend\,\,Effect} \nonumber \\
\Big(\frac{\rho_{\star,\obs}}{\rho_{\star,\tru}}\Big)^{\mathrm{PB}} &= \mathcal{B}^{-3/4} \Bigg(\frac{(1+\sqrt{\mathcal{B}} p_{\obs})^2-b_{\obs}^2}{(1+p_{\obs})^2-b_{\obs}^2}\Bigg)^{3/2}.
\label{eqn:photoblend}
\end{align}

Equation~\ref{eqn:photoblend} is maximized for $b_{\obs}\to (1-p_{\obs})$ and
$p_{\tru}\to1$ and for a binary star scenario of $\mathcal{B}\simeq2$, we
estimate that the PB effect cause AP effects up to order-unity. 
The assumption made to derive Equation~\ref{eqn:photoblend}, $(a/R_{\star})^2 
\gg (1+p)^2$, may also be expressed as

\begin{align}
%\mathbf{The\,\,}&\mathbf{Analytic\,\,Photoblend\,\,Condition} \nonumber \\
&\Big(\frac{P}{\mathrm{days}}\Big)^{4/3} \gg 0.389\,\Big(\frac{\rho_{\star,\tru}}{\mathrm{g\,cm}^3}\Big)^{-2/3}.
\label{eqn:photoblendcondition}
\end{align}

It can be seen from the above that this condition should be satisfied for all
but the very shortest of orbital periods (e.g. Kepler-78b; 
\citealt{sanchis:2013}). Since all blend sources must satisfy $\mathcal{B}>1$ 
(there is no such thing as a negative flux source), then inspection of 
Equation~\ref{eqn:photoblend} reveals that blends always cause one to 
underestimate the stellar density. In principle then, an independent measure of 
the stellar density can be used to measure the blend factor $\mathcal{B}$ by
inverting Equation~\ref{eqn:photoblend}. As discussed in detail in
Appendix~\ref{sub:blendinversion}, inverting Equation~\ref{eqn:photoblend}
yields a quadratic equation with two valid roots:

\begin{align}
\mathcal{B}_{+,-} =& \frac{1}{4p_{\obs}^4} \Bigg( -2p_{\obs} + \Big(\frac{\rho_{\star,\obs}}{\rho_{\star,\tru}}\Big)^{2/3} [(1+p_{\obs})^2-b_{\obs}^2] \nonumber\\
\qquad& \pm \Bigg[ \Big( p_{\obs} \Big[(2+p_{\obs})\Big(\frac{\rho_{\star,\obs}}{\rho_{\star,\tru}}\Big)^{2/3}-2\Big] \nonumber\\
\qquad& + \Big(\frac{\rho_{\star,\obs}}{\rho_{\star,\tru}}\Big)^{2/3} (1-b_{\obs}^2) \Big)^2 - 4 p_{\obs}^2 (1-b_{\obs}^2) \Bigg]^{1/2} \Bigg).
\label{eqn:Bplusminus}
\end{align}

\begin{figure*}
\begin{center}
\includegraphics[width=16.8 cm]{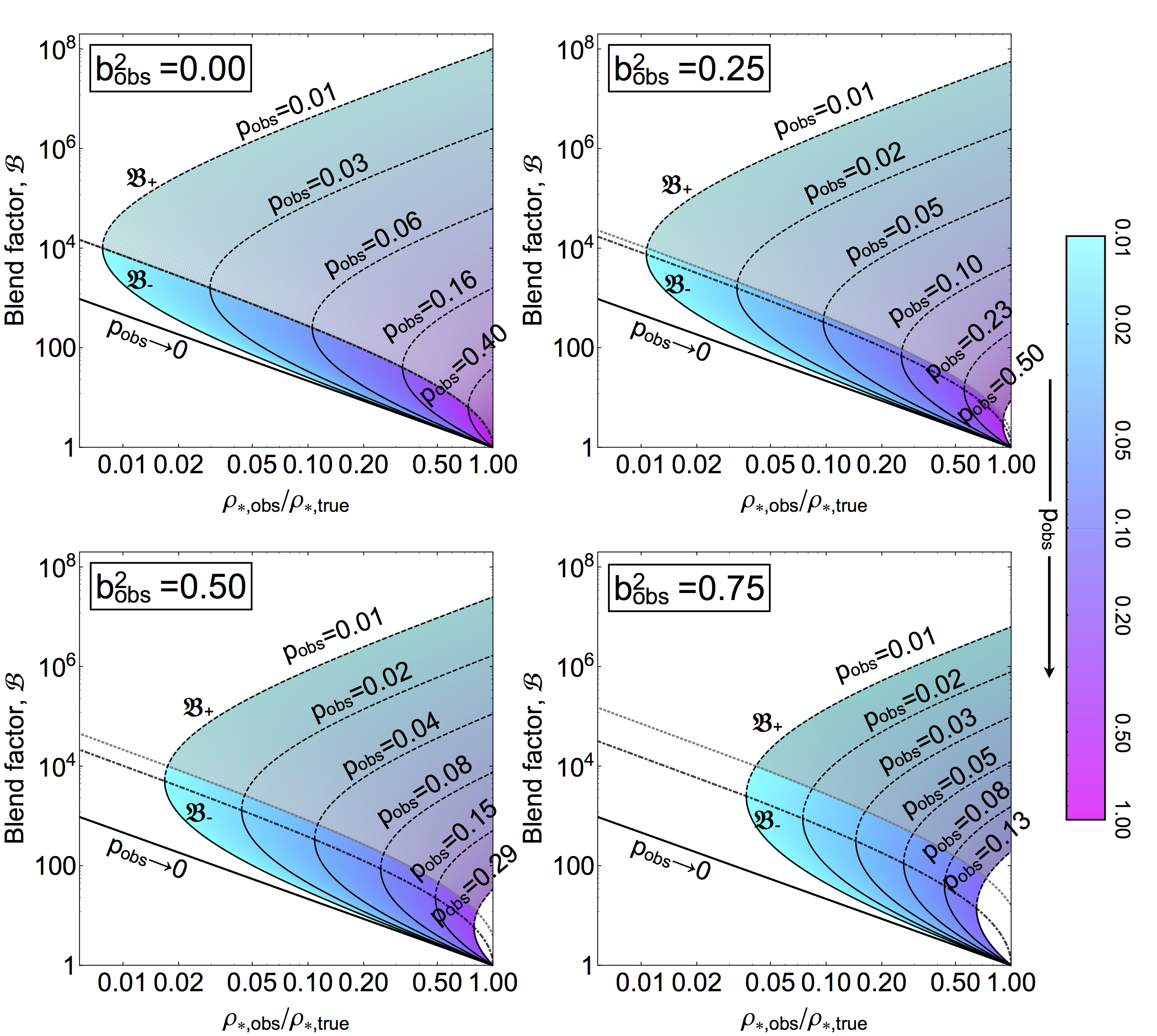}
\caption{\emph{\textbf{The Photo-blend Effect:} Blends, or uncorrected 
contaminated light, always cause one to underestimate the stellar density,
plotted here on the $x$-axis as $(\rho_{\star,\obs}/\rho_{\star,\tru})$. One 
may solve for the blend factor, $\mathcal{B}$, to aid in validating candidate 
planets, yielding two analytic roots shown by the curved black 
($\mathcal{B}_{+}$) and black-dashed ($\mathcal{B}_{-}$) lines for a range of 
apparent impact parameters, $b_{\obs}$, and ratio-of-radii, $p_{\obs}$. The 
$\mathcal{B}_{+}$ root is only physically valid between the point of inflection 
of the contours (traced by the black dot-dashed line) and the dotted gray 
line.}} 
\label{fig:Bcontours}
\end{center}
\end{figure*}

The $\mathcal{B}_{+,-}$ functions are plotted in Figure~\ref{fig:Bcontours}
for different input parameters. There are several key observations of the
expression. Firstly, for a known $p_{\obs}$ and $b_{\obs}$,
$(\rho_{\star,\obs}/\rho_{\star,\tru})$ is always bound by the range:

\begin{align}
\Big(\frac{\rho_{\star,\obs}}{\rho_{\star,\tru}}\Big)_{\mathrm{min}} \leq \Big(\frac{\rho_{\star,\obs}}{\rho_{\star,\tru}}\Big) < \Big(\frac{\rho_{\star,\obs}}{\rho_{\star,\tru}}\Big)_{\mathrm{max}},
\end{align}

where

\begin{align}
&\Big(\frac{\rho_{\star,\obs}}{\rho_{\star,\tru}}\Big)_{\mathrm{min}} = \Bigg(
\frac{2 p_{\obs} (1+\sqrt{1-b_{\obs}^2})}{ (1+p_{\obs})^2 - b_{\obs}^2 } \Bigg)^{3/2},\\
&\Big(\frac{\rho_{\star,\obs}}{\rho_{\star,\tru}}\Big)_{\mathrm{max}} = 1.
\label{eqn:rholimits}
\end{align}

Curiously then, there is both an upper and lower limit on the range of
$\rho_{\star,\obs}$ values a blend can produce any observation outside of this
range \emph{cannot} be due to the photo-blend effect only.

The second important observation is although the solution for $\mathcal{B}$ is
bi-modal, it is actually uni-modal for most 
$(\rho_{\star,\obs}/\rho_{\star,\tru})$ inputs. Specifically, as shown in
Appendix~\ref{sub:blendinversion}, the $\mathcal{B}_{+}$ is unphysical most
inputs. This is also illustrated in Figure~\ref{fig:Bcontours} by the gray
dotted line. In practice then, only a small range of parameter space is
bi-modal, which occurs when:

\begin{align}
\Big(\frac{\rho_{\star,\obs}}{\rho_{\star,\tru}}\Big)_{\mathrm{min}} \leq \Big(\frac{\rho_{\star,\obs}}{\rho_{\star,\tru}}\Big) < \Big(\frac{\rho_{\star,\obs}}{\rho_{\star,\tru}}\Big)_{\mathcal{B}_{+},\mathrm{max}},
\end{align}

where

\begin{align}
\Big(\frac{\rho_{\star,\obs}}{\rho_{\star,\tru}}\Big)_{\mathcal{B}_{+,\mathrm{max}}} &= \Bigg( \frac{ (4-b_{\obs}^2) p_{\obs} }{ (1+p_{\obs})^2-b_{\obs}^2 } \Bigg)^{3/2}.
\end{align}

In fact, as visible in Figure~\ref{fig:Bcontours}, this bi-modal range has
zero volume as $b_{\obs}\rightarrow0$ since
$\big(\frac{\rho_{\star,\obs}}{\rho_{\star,\tru}}\big)_{\mathrm{min}}
=\big(\frac{\rho_{\star,\obs}}{\rho_{\star,\tru}}\big)_{\mathcal{B}_{+},
\mathrm{crit}}$ in this limit. In summary then, we have:

\begin{equation}
\mathcal{B} =
\begin{cases}
\mathrm{no\,\,roots}  & \text{if } 0<\Big(\frac{\rho_{\star,\obs}}{\rho_{\star,\tru}}\Big)<\Big(\frac{\rho_{\star,\obs}}{\rho_{\star,\tru}}\Big)_{\mathrm{min}} \\
\mathcal{B}_{-}\,\,\mathrm{or}\,\,\mathcal{B}_{+} & \text{if } \Big(\frac{\rho_{\star,\obs}}{\rho_{\star,\tru}}\Big)_{\mathrm{min}}<\Big(\frac{\rho_{\star,\obs}}{\rho_{\star,\tru}}\Big)<\Big(\frac{\rho_{\star,\obs}}{\rho_{\star,\tru}}\Big)_{\mathcal{B}_{+},\mathrm{crit}} \\
\mathcal{B}_{-}  & \text{if } \Big(\frac{\rho_{\star,\obs}}{\rho_{\star,\tru}}\Big)_{\mathcal{B}_{+},\mathrm{crit}}<\Big(\frac{\rho_{\star,\obs}}{\rho_{\star,\tru}}\Big)<1 \\
\mathrm{no\,\,roots}  & \text{if } 1<\Big(\frac{\rho_{\star,\obs}}{\rho_{\star,\tru}}\Big)<\infty
\end{cases}
\end{equation}

\subsection{The Photo-spot Effect}
\label{sub:photospot}

Starspots, networks and plages are thought to form by stellar magnetic fields
generated by cyclonic turbulence in the outer convection zone of cool stars
penetrating the stellar atmosphere \citep{berdyugina:2005}. Starspots are
thought to be a particularly common outcome of this process and continuous 
photometric monitoring reveals their signature as rotational modulations, which 
has allowed for the determination of rotation periods for thousands of stars 
\citep{basri:2011,walkowicz:2013}.

Whilst large spots which are occulted by the transiting object are easy to
identify and remove, unocculted spots are more challenging and perturb the
transit depth as pointed out by \citet{czesla:2009}. We define the act of 
unocculted starspots perturbing the observed transit depth, and thus the
observed stellar density, as the ``photo-spot'' effect.

Equation~\ref{eqn:photoblend} reveals that since $\mathcal{B}\geq 1$ for all
blend sources, then the effect of a blend is to underestimate the stellar 
density. However, as proved in \citet{macula:2012}, the transit depth change due 
to unocculted starspots behave like a $\mathcal{B}<1$ blend factor and actually 
enhance the transit depth. This would therefore cause an observer to measure 
$(\rho_{\star,\obs}/\rho_{\star,\tru})>1$. \citet{macula:2012} showed that the 
effect of the transit depth is given by

\begin{align}
\frac{\delta_{\obs}}{\delta_{\tru}} &= \frac{F_{\star}(\mathrm{unspotted})}{F_{\star}(\mathrm{spotted})},\nonumber\\
\frac{\delta_{\obs}}{\delta_{\tru}} &= \frac{ 1 }{ 1 - A_{\mathrm{spots}} },
\end{align}

where $F_{\star}$ is the flux from the star and the unspotted case corresponds 
to the flux an observer would see if one took the actual starspot population and 
shrunk their sizes to zero. The second line re-writes this expression by
defining $A_{\mathrm{spots}}$ as the effective normalized photometric 
amplitude of the rotational modulations. For a rotating star with one to a few 
major spots, there will be times when all of the spots in view and times when no 
spots are present, giving rise to quasi-periodic transit depth variations 
(T$\delta$V). We assume such a rotation period a) much longer than the transit
duration, b) much shorter than the baseline of observations and c) has no
commensurability with the transiting body's orbital period. If we treat 
$F_{\star}(\mathrm{spotted})$ as behaving like a Fourier series of harmonic 
components, then the average effect on the transit depths (i.e. the folded 
transit light curve depth) would be

\begin{align}
\frac{\bar{\delta}_{\obs}}{\delta_{\tru}} &\simeq \frac{ 1 }{ 1 - A_{\mathrm{spots}}/2 },
\end{align}

\begin{figure*}
\begin{center}
\includegraphics[width=16.8 cm]{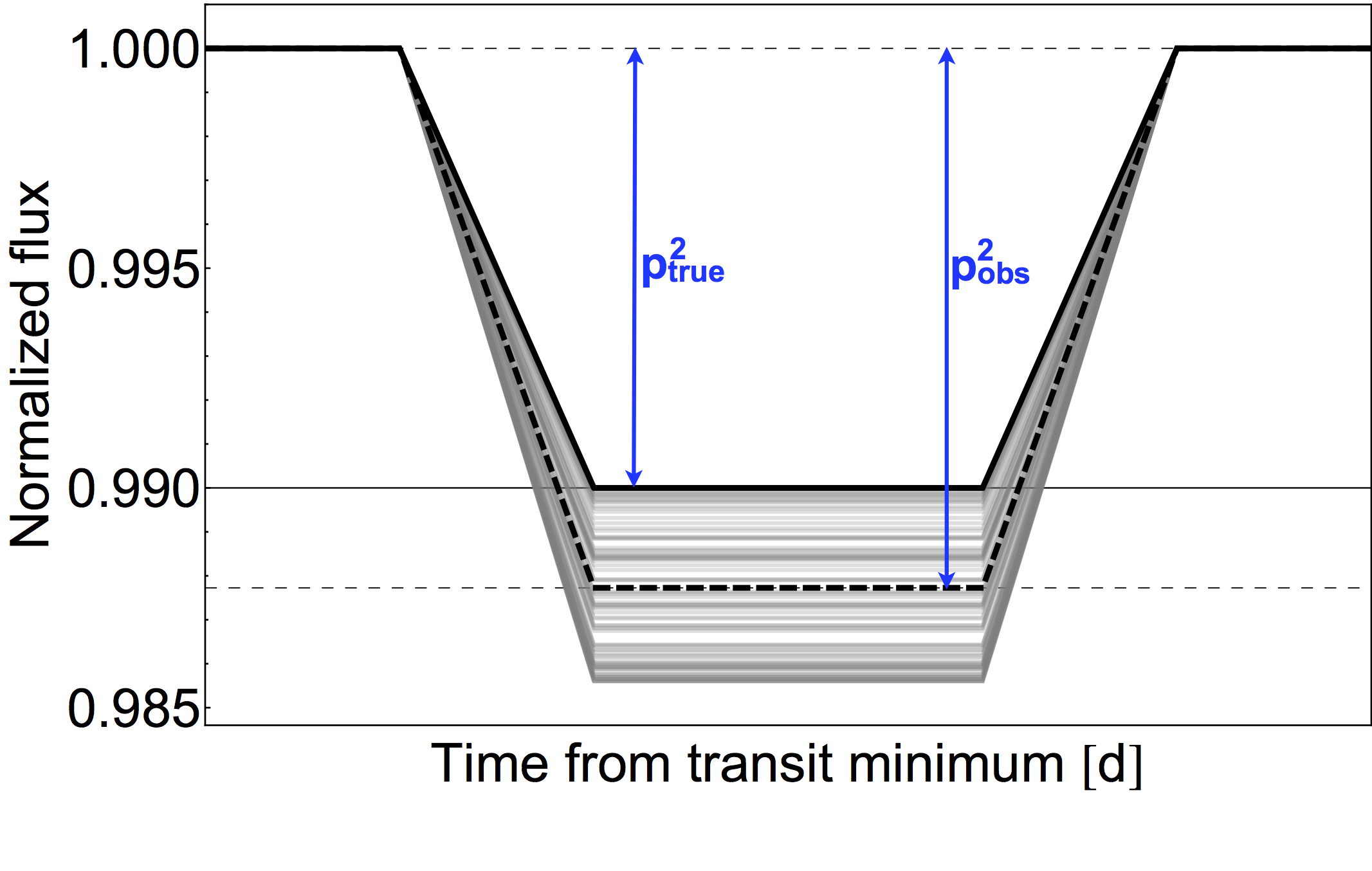}
\caption{\emph{\textbf{The Photo-spot Effect:} By neglecting to correct for 
transit depth variations (T$\delta$V) due to unocculted spots, a folded transit 
light curve will exhibit deformation leading to the erroneous retrieval of the 
basic transit parameters, including the observed stellar density, 
$\rho_{\star,\obs}$. Here, the black line represents the true original signal, 
the gray lines are 100 examples of the signal with unaccounted for sinusoidal 
T$\delta$Vs and the black-dashed line is the naively folded transit light curve, 
exhibiting sizable deformation.
}} 
\label{fig:spotplot}
\end{center}
\end{figure*}

The photo-spot effect is illustrated in Figure~\ref{fig:spotplot}, where the 
T$\delta$Vs give rise an apparently increased depth in the folded light curve.
The depth ratio, $(\bar{\delta}_{\obs}/\delta_{\tru})$, is equivalent to 
$\mathcal{B}^{-1}$ using our definition of the blend factor in 
Equation~\ref{eqn:Beqn}. Exploiting this trick, one may write that a spot
behaves like a blend with a blend factor, $\mathcal{B}_{\mathrm{spot}}$, given
by

\begin{align}
\mathcal{B}_{\mathrm{spot}} &\simeq 1 - \frac{A_{\mathrm{spots}}}{2}, \nonumber\\
\mathcal{B}_{\mathrm{spot}} &\simeq \frac{1}{2} \Bigg( 1 + \frac{F_{\star}(\mathrm{spotted})}{F_{\star}(\mathrm{unspotted})}\Bigg).
\label{eqn:Bspot}
\end{align}

Equipped with Equation~\ref{eqn:Bspot}, one may now compute the consequences
on the stellar density using the same expressions derived earlier for the
photo-blend effect in \S\ref{sub:photoblend}. 

\begin{align}
%&\mathbf{The\,\,}\mathbf{Analytic\,\,Photospot\,\,Effect} \nonumber \\
&\Big(\frac{\rho_{\star,\obs}}{\rho_{\star,\tru}}\Big)^{\mathrm{PS}} = \lim_{\mathcal{B}\to\mathcal{B}_{\mathrm{spot}}} \Big(\frac{\rho_{\star,\obs}}{\rho_{\star,\tru}}\Big)^{\mathrm{PB}}
\label{eqn:photospot}
\end{align}

We note that plotting the $\mathcal{B}_{+}$ root for 
$(\rho_{\star,\obs}/\rho_{\star,\tru})>1$ again yields unphysically high blend 
factors (and always $\mathcal{B}>1$ which is not possible from the photo-spot 
effect). Therefore, one only need consider the $\mathcal{B}_{-}$ root for the 
photo-spot effect. As with the photo-blend effect, the same conditions apply
for the application of these analytic photo-spot equations:

\begin{align}
%\mathbf{The\,\,}&\mathbf{Analytic\,\,Photospot\,\,Condition} \nonumber \\
&\Big(\frac{P}{\mathrm{days}}\Big)^{4/3} \gg 0.389\,\Big(\frac{\rho_{\star,\tru}}{\mathrm{g\,cm}^3}\Big)^{-2/3}.
\label{eqn:photospotcondition}
\end{align}

%Accordingly, the spot amplitude, $A_{\mathrm{spot}}$ is recoverable via:
%
%\begin{align}
%A_{\mathrm{spot}} &= 2 (1-\mathcal{B}_{-}) \\
%\end{align}

Typically, even a heavily spotted star will be in the range 
$A_{\mathrm{spots}}\lesssim20$\% and usually $\lesssim1$\%. Therefore
spots affect the ratio $(\rho_{\star,\obs}/\rho_{\star,\tru})$ at the same
order-of-magnitude level as the normalized rotational modulations amplitude.
The maximum AP deviation via this effect can be evaluated by computing the
limit for $b_{\obs}\to(1-p_{\obs})$ and $p_{\tru}\to1$. For an extreme 20\% spot
amplitude we obtain an AP effect of order $\mathcal{O}[10^{-1}]$, and for a 
typical 1\% spot amplitude this becomes $\mathcal{O}[10^{-2}]$.
In principle, it is possible to correct for the photo-spot effect using 
rotational modulation data, although this can be challenging \citep{macula:2012} 
and such effort should be put in the context of the expected magnitude of this 
effect.

\subsection{The Photo-timing Effect}
\label{sub:phototiming}

\begin{figure*}
\begin{center}
\includegraphics[width=16.8 cm]{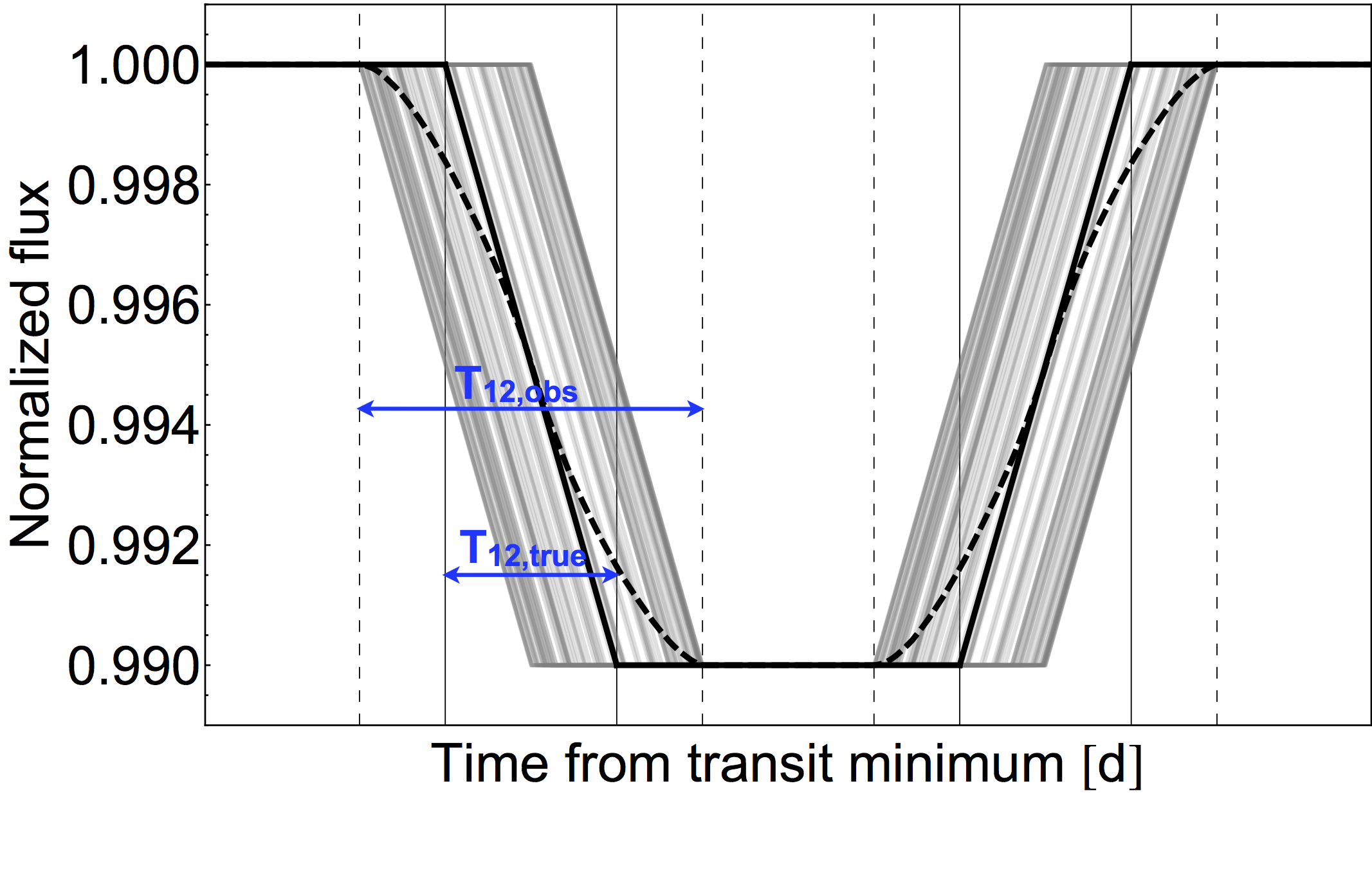}
\caption{\emph{\textbf{The Photo-timing Effect:} If even low-amplitude transit 
timing variations (TTV) are negated, a folded transit light curve will exhibit
deformation leading to the erroneous retrieval of the basic transit parameters,
including the observed stellar density, $\rho_{\star,\obs}$. Here, the black 
line represents the true original signal, the gray lines are 100 examples of the
signal with unaccounted for sinusoidal TTVs and the black-dashed line is the
naively folded transit light curve, exhibiting sizable deformation.
}} 
\label{fig:TTVplot}
\end{center}
\end{figure*}

Transit timing variations (TTVs) have been revealed by the \emph{Kepler Mission}
to be a fairly common occurrence in planetary systems 
\citep{ford:2012,mazeh:2013} with $\sim$10\% showing significant TTVs. TTVs of
low amplitude can be difficult to infer by fitting individual transits and yet
if we ignore their presence they will systematically bias the derived transit
parameters. An object with low-amplitude TTVs ($A_{\mathrm{TTV}}<T_{23}$)
with $N\gg1$ cycles over the baseline of continuous transit observations will
cause a naively folded transit light curve to appear smeared out, as illustrated 
in Figure~\ref{fig:TTVplot}. The four contact points appear shifted due to the
motion of the planet leading to erroneously derived $T_{23}$ and $T_{14}$
durations. Naturally, this will feed into the derived impact parameter,
scaled semi-major axis and light curve derived stellar density.

In Appendix~\ref{app:phototiming}, we derive the full consequences of unaccounted
TTVs on the derived transit parameters. The effect on $\rho_{\star}$ depends
upon the true impact parameter but unfortunately the impact parameter is
also corrupted by the TTVs. One way round this is to consider the worst-case
scenario where $\rho_{\star,\obs}$ is most discrepant from $\rho_{\star,\tru}$,
which occurs for $b=0$. In this case, one finds a simple form for the
photo-timing effect:

\begin{align}
%&\mathbf{The\,\,}\mathbf{Analytic\,\,Phototiming\,\,Effect} \nonumber \\
&\Big( \frac{ \rho_{\star,\obs} }{ \rho_{\star,\tru} }\Big)^{\mathrm{PT}} \geq \Bigg( \frac{p}{p+n A_{\mathrm{TTV}} (a/R_{\star})} \Bigg)^{3/2},
\label{eqn:TTVeffect}
\end{align}

where $n=2\pi/P$, $2A_{\mathrm{TTV}}$ is the peak-to-peak TTV amplitude,
$(a/R_{\star})$ implicitly refers to $(a/R_{\star})_{\tru}$
and we use the $\geq$ symbol since the calculation is computed for the extreme
case of $b=0$. In Appendix~\ref{app:phototiming}, we show that this expression
is a valid approximation for:

\begin{align}
%&\mathbf{The\,\,Analytic\,\,Phototiming\,\,Condition} \nonumber \\
&(a/R_{\star})^2 \gg 2,\\
&2A_{\mathrm{TTV}} \ll T_{23}.
\end{align}

The first condition may also be re-expressed in physical units as

\begin{align}
&\Big(\frac{P}{\mathrm{days}}\Big)^{4/3} \gg 0.231\,\Big(\frac{\rho_{\star,\tru}}{\mathrm{g\,cm}^3}\Big)^{-2/3}.
\end{align}

In general, one expects TTVs to be detectable by careful 
inspection of the data. In the case that significant TTVs are present, an 
accurate light curve derived stellar density could be derived by either using a 
model which allows for unique transit times or using a physical model which 
accounts for TTVs (i.e. a photodynamical model), provided the physical model 
well-explains the data.

If no significant TTVs are detected, or an observer opts to try and remove the
best-fitting TTVs and then re-fit assuming a linear ephemeris, the light curve
derived stellar density can still be affected by unseen low-amplitude TTVs.
In principle, one expects to be able to exclude TTVs up to some maximum
amplitude level to 1-, 2-, 3- (etc) $\sigma$ confidence. In essence, this means
that the uncertainty on $\rho_{\star,\obs}$ will be underestimated. However,
using our expressions, it is possible to quantify this unaccounted-for
uncertainty in the extreme case occurring for $b_{\tru}=0$:

\begin{align}
\sigma_{(\rho_{\star,\obs}/\rho_{\star,\tru})} &\lesssim 7.5 \frac{G^{1/3}\rho_{\star,\obs}^{1/3}}{p P^{1/3}} \frac{\sigma_{\tau}}{N^{1/4}},
\label{eqn:TTVerror}
\end{align}

where $\sigma_{\tau}$ is the typical timing uncertainty on each transit and
$N$ is the number of transits observed. As an example, consider a planet with 
$P=10$\,days, $p=0.1$ around a Solar-like star. Consider that each transit can 
be timed to a precision of 1\,minute and that over a span of 4\,years the target 
is continuously monitored. This would give 
$\sigma_{(\rho_{\star,\obs}/\rho_{\star,\tru})}^{\mathrm{max}}\lesssim9\%$.
Once again, we emphasize that this error would not be normally propagated into
the uncertainty on $\rho_{\star,\obs}$. This demonstrates the photo-timing
effect leads to inflated errors on the stellar density and caution must be taken
in interpretting small discrepancies. For much larger timing errors of 
$\sigma_{\tau}\simeq10$\,minutes, the effect can be estimated to be
$\mathcal{O}[10^0]$.

\subsection{The Photo-duration Effect}
\label{sub:photoduration}

Transit durations variations (TDVs) are another example of a dynamical effect
which will alter the shape of a folded transit light curve, if left unaccounted
for. TDVs were first posited to be a signature of exomoons 
\citep{kipping:2009a,kipping:2009b} but have since been demonstrated to be
also possible in strongly interacting multi-planet systems too 
\citep{nesvorny:2012}. TDVs come in two flavors, velocity-induced transit
duration variations, TDV-V, and transit impact parameter induced transit
duration variations, TDV-TIP \citep{kipping:2009a,kipping:2009b}. In this work,
we focus on the more dominant component of TDV-V.

\begin{figure*}
\begin{center}
\includegraphics[width=16.8 cm]{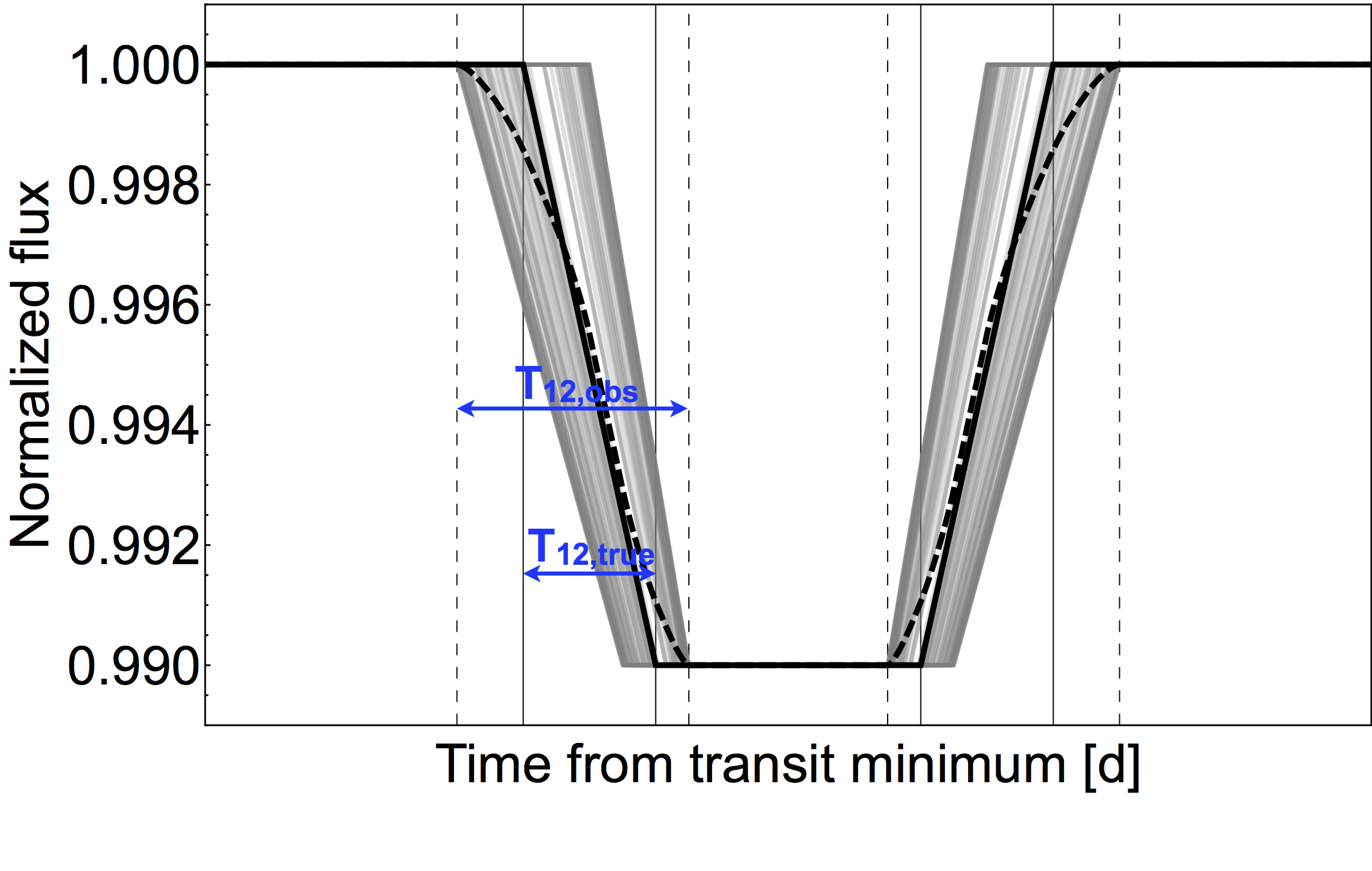}
\caption{\emph{\textbf{The Photo-duration Effect:} If even low-amplitude transit 
duration variations (TDVs) are negated, a folded transit light curve will 
exhibit deformation leading to the erroneous retrieval of the basic transit 
parameters, including the observed stellar density, $\rho_{\star,\obs}$. Here, 
the black line represents the true original signal, the gray lines are 100 
examples of the signal with unaccounted for sinusoidal TDVs and the black-dashed 
line is the naively folded transit light curve, exhibiting sizable deformation.
}} 
\label{fig:TDVplot}
\end{center}
\end{figure*}

TDV-Vs essentially stretch and squash the width of the transit shape and a well
sampled periodic set of light curves with TDV-Vs will exhibit a deformed folded
transit shape, if neglected. This is illustrated in Figure~\ref{fig:TDVplot},
where the composite light curve is deformed in a similar way to that caused
by periodic TTVs earlier in Figure~\ref{fig:TTVplot}. We consider the TDVs
to be due to the velocity of a planet varying periodically between the
extrema of $v_{\mathrm{min}}=v_0 (1-A_{\mathrm{TDV}})$ and 
$v_{\mathrm{max}}=v_0 (1+A_{\mathrm{TDV}})$, where the ``0'' subscript denotes
the parameter's value in the absence of TDVs. Since the durations are inversely
proportional to the velocity of the planet, then the durations vary over time
$t$ over the range:

\begin{align}
T_{ _{23}^{14} ,0} (1-A_{\mathrm{TDV}})\leq T_{ _{23}^{14} }(t) \leq T_{ _{23}^{14} ,0} (1+A_{\mathrm{TDV}}).
\end{align}

The $A_{\mathrm{TDV}}$ term therefore defines the relative changes in the
duration, and not the absolute changes, which is the more natural expression of
TDV-Vs. Using this model, we derive the effect of periodic TDVs on the light 
curve derived stellar density in Appendix~\ref{app:photoduration} to be:

\begin{align}
%&\mathbf{The\,\,}\mathbf{Analytic\,\,Photoduration\,\,Effect} \nonumber \\
&\Big(\frac{\rho_{\star,\obs}}{\rho_{\star,\tru}}\Big)^{\mathrm{PD}} = \nonumber\\
&\Bigg( \frac{ (a/R_{\star})^2 p + 4 A_{\mathrm{TDV}}^2 b^2 p + 2 A_{\mathrm{TDV}} [ (1-p^2)^2-b^2(1+p^2) ] }{ (a/R_{\star})^2 [p+4A_{\mathrm{TDV}}^2p+2A_{\mathrm{TDV}} (1+p^2-b^2)] } \Bigg)^{3/2},
\label{eqn:photodurationapprox}
\end{align}

where $(a/R_{\star})$ is $(a/R_{\star})_{\tru}$ and can be estimated as 
$[(G P^2 \rho_{\star})/(3\pi)]^{1/3}$. In Appendix~\ref{app:photoduration}, we
show that the above is valid when

\begin{align}
%&\mathbf{The\,\,Analytic\,\,Photoduration\,\,Condition} \nonumber \\
&(a/R_{\star})^2 \gg 2,\\
&A_{\mathrm{TDV}} \ll 1.
\end{align}

As with the photo-timing effect, the photo-duration effect can be thought of
as imparting an error term on the observed stellar density. We are unable to
find a simple form for the resulting expression though and so suggest observers
use:

\begin{align}
\sigma_{(\rho_{\star,\obs}/\rho_{\star,\tru})}^{\mathrm{PD}} &= 1 - \Bigg[ \lim_{ A_{\mathrm{TDV}} \to\sigma_{A_{\mathrm{TDV}}} } \Bigg(\frac{\rho_{\star,\obs}}{\rho_{\star,\tru}}\Bigg)^{\mathrm{PD}} \Bigg].
\label{eqn:TDVerror}
\end{align}

As with the photo-timing effect error, we demonstrate the above by considering
the same example of a planet with $P=10$\,days, $p=0.1$ around a Solar-like 
star with $b=0$ ($T_{14,0}=4.3$\,hours). Timing each transit to 1\,minute 
precision, which corresponds to an approximately 2\,minute duration uncertainty, 
over 4\,years of continuously monitoring gives 
$\sigma_{(\rho_{\star,\obs}/\rho_{\star,\tru})}^{\mathrm{max}}=10.6\%$.
Therefore, in this example, the photo-duration effect imparts approximately the 
same level of uncertainty into the observed stellar density as the photo-timing 
effect does. However, unlike TTVs, TDVs are considerably rarer in the database
of known exoplanets with only a few examples and so the a-priori probability
of hidden TDVs is clearly distinct to that from timing variations.

\subsection{The Photo-eccentric Effect}
\label{sub:photoeccentric}

\subsubsection{General effect}

The effect of eccentricity is the most well-studied asterodensity profiling
effect. \citet{dawson:2012a} refer to this asterodensity profiling effect as
the ``photo-eccentric effect'', as we do so here.  The first explicit derivation 
of the effect of eccentricity is given in \citet{map:2012} who find

\begin{align}
%\mathbf{The\,\,Analytic\,\,}&\mathbf{Photoeccentric\,\,Effect} \nonumber \\
&\Big( \frac{\rho_{\obs}}{\rho_{\tru}} \Big)^{\mathrm{PE}} = \Psi,
\label{eqn:eeffect}
\end{align}

where

\begin{align}
\Psi &\equiv \frac{(1+e\sin\omega)^{3}}{(1-e^2)^{3/2}}.
\end{align}

Despite the expression already existing in the literature, we are unaware of
any investigations regarding the range of parameters which the 
\emph{approximation} shown in Equation~\ref{eqn:eeffect} is valid. In 
Appendix~\ref{app:psi}, we present a detailed investigation of this and
surmise that the above is valid for:

\begin{align}
%\mathbf{The\,\,}&\mathbf{Analytic\,\,Photoeccentric\,\,Condition} \nonumber \\
&(a/R_{\star})^2 \gg \frac{2}{3} \Bigg( \frac{1+e}{1-e} \Bigg)^3,
\end{align}

which may also be expressed in physical units as

\begin{align}
\Bigg(\frac{P}{\mathrm{days}}\Bigg)^{4/3} \gg 0.101 \Bigg( \frac{\rho_{\star,\tru}}{\mathrm{g\,cm}^{-3}} \Bigg)^{-2/3} \Bigg( \frac{1+e}{1-e} \Bigg)^3.
\end{align}

As pointed out in numerous previous works \citep{burke:2008,kipping:2008,
winn:2010,dawson:2012a}, if a planet on an eccentric orbit is observed to 
transit, then a-priori it is more probable that $0<\omega\leq\pi$ than 
$\pi<\omega\leq2\pi$. This is because the geometric transit probability is given 
by

\begin{align}
\mathrm{P}(b\leq1) &= \frac{1}{(a/R_{\star})} \frac{1+e\sin\omega}{1-e^2}
\end{align}

and so

\begin{align}
\frac{\mathrm{P}(0<\omega\leq\pi|b\leq1)}{\mathrm{P}(\pi<\omega\leq2\pi|b\leq1)} &= \frac{\pi+2e}{\pi-2e}
\end{align}

which is greater than 1 for all $0<e\leq1$. Note that the exact ratio cannot be 
estimated without assuming some prior distribution for the eccentricity.
The consequence of this is that $\Psi>1$ is a more probable result than $\Psi<1$ 
by the odds-ratio derived above, given that a planet is transiting and 
eccentric. This is an important result because it means eccentric orbits 
\emph{tend} to overestimate $(\rho_{\star,\obs}/\rho_{\star,\tru})$ whereas all 
of the previous effects discussed, except the photo-mass effect, underestimate 
the density. However, it is also worth noting that even for moderately high
eccentricities of say $e\sim0.5$, the odds-ratio quoted above is $\sim2$ and
thus although an overestimated density is more likely from the photo-eccentric
effect, it is not dramatically so.

\subsubsection{Minimum eccentricity}

The photo-eccentric effect directly reveals $\Psi$, which is a function of
both $e$ and $\omega$. Ideally, one wishes to obtain information on both
$e$ and $\omega$ in isolation, but purely from an information theory perspective
it is obvious this ideal can never be truly realized, since we have one
measurement and two unknowns. Progress can be made by considering the minimum
eccentricity. In the Appendix~\ref{app:photoeccentric}, we show that the
\emph{minimum} eccentricity can be derived in the case of SAP and provide a
single-domain function for $e_{\mathrm{min}}$ as

%\begin{align}
%e_{\min} =& \Big( \frac{\Psi^{2/3}-1}{\Psi^{2/3}+1} \Big) \mathbb{H}[\Psi-1] \nonumber \\
%\quad& + \Big( \frac{ (1-\Psi^{2/3})(1-\Psi^{2/3}+\Psi^{4/3}) }{1+\Psi^2} \Big) \mathbb{H}[-\Psi+1],
%\label{eqn:emin}
%\end{align}

\begin{align}
e_{\min} =& \Bigg( \frac{ \big(\frac{\rho_{\star,\obs}}{\rho_{\star,\tru}}\big)^{2/3}-1}{\big(\frac{\rho_{\star,\obs}}{\rho_{\star,\tru}}\big)^{2/3}+1} \Bigg) \mathbb{H}\Big[\Big(\frac{\rho_{\star,\obs}}{\rho_{\star,\tru}}\Big)-1\Big] \nonumber \\
\quad& + \Bigg( \frac{ (1-\big(\frac{\rho_{\star,\obs}}{\rho_{\star,\tru}}\big)^{2/3})(1-\big(\frac{\rho_{\star,\obs}}{\rho_{\star,\tru}}\big)^{2/3}+\big(\frac{\rho_{\star,\obs}}{\rho_{\star,\tru}}\big)^{4/3}) }{1+\big(\frac{\rho_{\star,\obs}}{\rho_{\star,\tru}}\big)^2} \Bigg) \nonumber\\
\qquad& \mathbb{H}\Big[1-\Big(\frac{\rho_{\star,\obs}}{\rho_{\star,\tru}}\Big)\Big],
\label{eqn:emin}
\end{align}

where $\mathbb{H}[x]$ is the Heaviside Theta function. We note that previous
authors have derived or discuss double-domain functions for $e_{\mathrm{min}}$ 
such as \citet{barnes:2007} and \citet{kane:2012}. The single-domain function
presented here simply combines the two domains using Heaviside Theta functions
and uses stellar density as the observable rather than durations.
We also stress that $e_{\min}$ is purely a function of 
$(\rho_{\star,\obs}/\rho_{\star,\tru})$ and no other terms. It is 
therefore possible to analytically calculate the uncertainty on $e_{\min}$ using 
quadrature:

\begin{align}
\sigma_{e_{\min}} &= \frac{4}{3} \Big(\frac{\rho_{\star,\obs}}{\rho_{\star,\tru}}\Big)^{-1/3}\Bigg(1+\Big(\frac{\rho_{\star,\obs}}{\rho_{\star,\tru}}\Big)^{2/3}\Bigg)^{-2} \sigma_{ \rho_{\star,\obs}/\rho_{\star,\tru} },
\label{eqn:eminerror}
\end{align}

where $\sigma_{e_{\min}}$ and $\sigma_{\rho_{\star,\obs}/\rho_{\star,\tru}}$ are 
the uncertainties on the minimum eccentricity and ratio of the observed to
true stellar density respectively.

\begin{figure*}
\begin{center}
\includegraphics[width=16.8 cm]{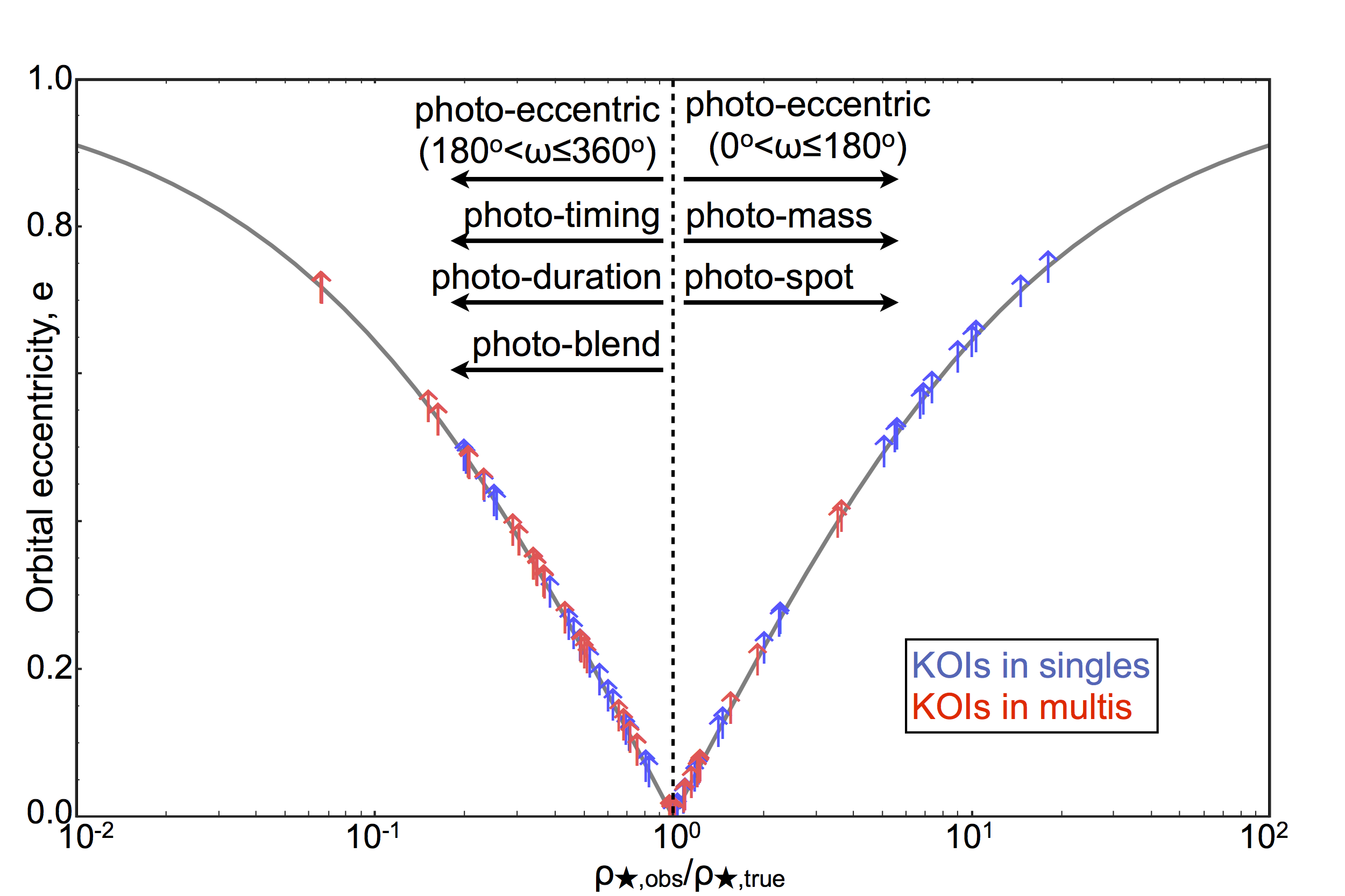}
\caption{\emph{\textbf{The Photo-eccentric Effect:} The minimum orbital 
eccentricity function, defined in Equation~\ref{eqn:emin}, plotted with respect 
to its only dependent variable, $(\rho_{\star,\obs},\rho_{\star,\tru})$. The 
arrows correspond to real KOIs with known asteroseismology measurements 
available, where blue are singles and red are multis. We also mark the 
directions in which the other asterodensity profiling effects act.
}} 
\label{fig:eminplot}
\end{center}
\end{figure*}

The simple $e_{\mathrm{min}}$ function is 
visualized in Figure~\ref{fig:eminplot}, where one can see the PE effect 
can induce AP deviations up to $\mathcal{O}[10^2]$. In this figure, we also 
over-plot Kepler Objects of Interest (KOIs) with asteroseismologically measured 
$\rho_{\star,\tru}$ values from \citet{huber:2013}. The $\rho_{\star,\obs}$ 
term is computed for each KOI using the MAST archival database
\footnote{http://archive.stsci.edu/kepler/koi/search.php} entries of 
$(a/R_{\star})$ and $P$. In principle, objects on the left-hand side (LHS) may 
be blends (since one does not expect a high proportion of eccentric planets on 
this side) and objects on the right are genuinely eccentric KOIs (therefore 
assuming that the photo-timing, photo-duration, photo-spot and photo-mass 
effects are minor). However, we caution that the large number of multis on the 
LHS, suggesting false-positive blended systems, is highly inconsistent with the 
expected low false-positive rates of multi-planet systems \citep{lissauer:2012}. 
We therefore advocate independent checks of these $\rho_{\star,\obs}$ values 
before drawing any conclusions, which is outside the scope of this work.

\subsubsection{Comparison to marginalization}

Calculating the minimum eccentricity using Equation~\ref{eqn:emin} is distinct 
from the strategy on the photo-eccentric effect by \citet{dawson:2012a} and 
\citet{dawson:2012b} who propose marginalizing over $\omega$, much like a 
nuisance parameter. The major advantage over marginalizing over $\omega$ is that 
one naturally incorporates the geometric transit probability effect and derives 
a singular estimate for $e$. This is useful since $e$ represents the most
physically useful parameter with respect to formation/evolution models
\citep{ford:2008,juric:2008,socrates:2012,dong:2013}.

However, there are several drawbacks of this approach compared to simply
computing $e_{\mathrm{min}}$ using Equation~\ref{eqn:emin}. Firstly, one can
only achieve this feat by assuming an a-priori distribution for the eccentricity
since the geometric transit probability is functionally dependent on both
$e$ and $\omega$, as discussed earlier. Therefore, the derived $e$ value
is fundamentally dependent upon the assumed prior distribution for $e$, which is
somewhat circular logic. In practice, \citet{dawson:2012a} found varying the 
priors on $e$ imposes only small changes in the derived posterior distributions 
of $e$, yet this we predict that this is only likely true where the data 
overwhelms the priors such as the cases considered by \citet{dawson:2012a} of 
highly eccentric systems causing $(\rho_{\star,\obs}/\rho_{\star,\tru})\gg1$.

Secondly, the act of marginalizing over a parameter which has
not been significantly constrained by the data fundamentally reduces the
information content of the final determination. In other words, one makes the
final determination fuzzier. In essence, the marginalization in $e$ space
is over $e_{\mathrm{min}}<e<1$, since $e<1$ for all bound orbits. Irrespective
of any reasonable prior on $e$, this will cause the marginalized $e$ value to
lie somewhere inbetween these two extrema and lead to elevated error bars
relative to $e_{\mathrm{min}}$ to accommodate this marginalization. Systems with
a high $e_{\mathrm{min}}$ will therefore appear to provide relatively small
errors on the marginalized $e$, purely because there is ``less room'' between
$e_{\mathrm{min}}$ and unity. An example of this is evident with the high
$e_{\mathrm{min}}$ system of HD~17156b reported in \citet{dawson:2012a} 
($e_{\mathrm{min}}\sim0.6$) giving $e=0.71_{-0.09}^{+0.16}$ whereas the 
$e_{\mathrm{min}}\sim0$ system of Kepler-22b yields much broader uncertainties 
of $e=0.13_{-0.13}^{+0.36}$ \citep{kepler22:2013}, despite both being bright 
targets with asteroseismology and high-quality photometry. In general then, we 
advocate at least providing the community with both the constrained, 
prior-independent $e_{\mathrm{min}}$ term in addition to the lossy, marginalized 
$e$.

Thirdly, the simple analytic form of our expression for $e_{\mathrm{min}}$
makes it attractive for rapid calculation on hundreds/thousands of systems.
Transits may be fitted en-masse assuming a circular orbit and then 
$e_{\mathrm{min}}$ is easily computed without any tacit assumption on the
$e$ distribution. An alternative but equivalent approach would be to compile
a database of $T_{14,\obs}$, $T_{23,\obs}$ and $\delta_{\obs}$ from which
one can also proceed to compute $e_{\mathrm{min}}$. Additionally, provided 
one knows the uncertainties on $\rho_{\star,\obs}$ and $\rho_{\star,\tru}$, then 
the uncertainty on $e_{\mathrm{min}}$ is easily recovered in a single expression 
given by Equation~\ref{eqn:eminerror}, at least under the assumption of no other
AP effects. We propose that this would be an advantageous strategy for upcoming
transit survey missions, such as TESS.

\subsection{False-Positives}

Let us define a ``false-positive'' planetary candidate to be one which orbits
a different star to that for which we have an independent measure of the stellar
density. In such a case, then it should be obvious that the two density 
estimates need not agree and can be grossly different (see \citealt{sliski:2014}
for examples of such cases). The exact difference will depend upon the spectral 
types of the two stars and the flux ratios. In this scenario, we have no 
information on $\rho_{\star,\tru}$, since the independent measure corresponds to 
a different star. We also know that the star hosting the transiting body must be 
heavily blended and so the photo-blend effect must be acting. If we ignore the 
other AP effects, we may recall from Equation~\ref{eqn:rholimits} that a limit 
exists on the maximum AP deviation due to the PB effect:

\begin{align}
\Big(\frac{\rho_{\star,\obs}}{\rho_{\star,\tru}}\Big) \geq \Bigg(
\frac{2 p_{\obs} (1+\sqrt{1-b_{\obs}^2})}{ (1+p_{\obs})^2 - b_{\obs}^2 } \Bigg)^{3/2}.
\end{align}

This may be re-expressed to constrain the unknown quantity $\rho_{\star,\tru}$
via:

\begin{align}
\rho_{\star,\mathrm{alt}} \geq \rho_{\star,\obs} \Bigg(
\frac{ (1+p_{\obs})^2 - b_{\obs}^2 }{2 p_{\obs} (1+\sqrt{1-b_{\obs}^2})} \Bigg)^{3/2},
\label{eqn:rhoalt}
\end{align}

where we replace $\rho_{\star,\tru}$ with $\rho_{\star,\mathrm{alt}}$ to stress
that the transiting body is orbiting an alternative star. We do not refer to
this scenario as a ``photo-\emph{name}'' effect, since unlike the other cases no
independent information on the true stellar density is available. However,
Equation~\ref{eqn:rhoalt}, which is only valid in the absence of the PE, PT,
PD, PM and PS effects, may be of use to observers vetting planetary candidates.

\section{Discussion}

\subsection{Consequences for Measuring Eccentricities with SAP}
\label{sub:SAP}

\begin{figure*}
\begin{center}
\includegraphics[width=16.8 cm]{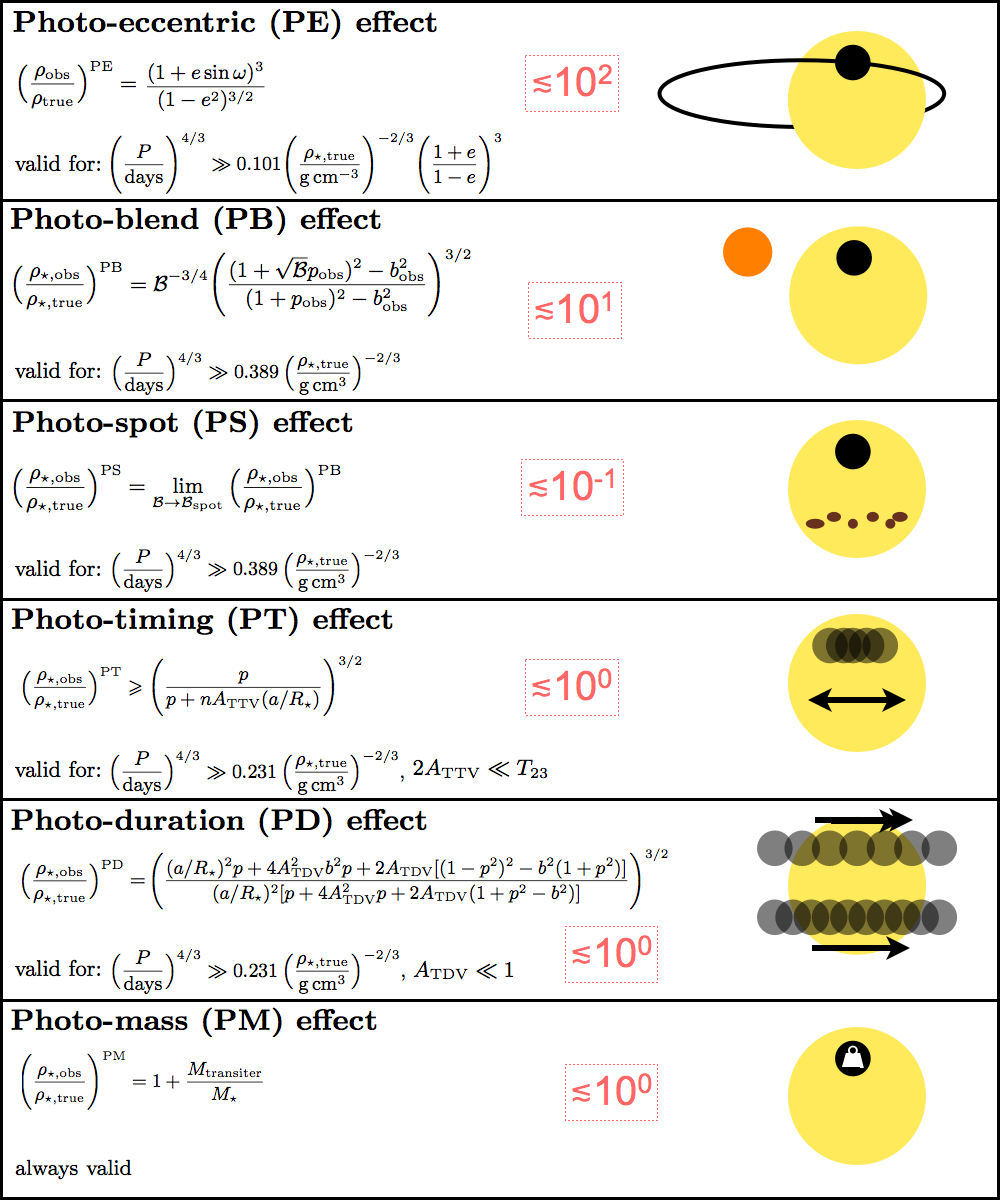}
\caption{\emph{\textbf{Asterodensity Profiling ``Cheat Sheet'':} Summary of
the analytic formulae for the various AP effects derived in this work,
including the supported parameter range for their applicability. Red 
boxes provide approximate order-of-magnitude for each effect. All effects
also assume $0<b<(1-p)$ i.e. a ``flat-bottomed'' transit.
}} 
\label{fig:summaryplot}
\end{center}
\end{figure*}

In this work, we have shown that at least five other asterodensity profiling
effects exist in addition to the photo-eccentric effect, which are
summarized in the ``cheat sheet'' of Figure~\ref{fig:summaryplot}. Since a 
number of phenomena can induce significant changes to the light curve derived 
stellar density, great care must be taken not to over-interpret any deviations 
as implying high eccentricity where none exists. Equivalently, one must be wary
of interpreting a lack of significant discrepancies as implying most planets
orbit on near-circular orbits. Put succinctly, the eccentricity distribution
can only be extracted using AP with a careful consideration of the prior
distributions for the other AP effects (e.g. the photo-blend effect, the 
photo-timing effect, etc). We would argue that any eccentricity distributions
purported without such due diligence cannot be considered physically 
representative of the true sample. This statement is justified by the fact
that not only are the other AP effects significant at the typical measurement
uncertainties, but they will impart systematic shifts to any naively computed
$e$ distribution, since the parameters upon which they depend are often skewed 
in one direction (e.g. for the photo-blend effect the blend parameter must 
always be $>1$ and thus always cause an underestimation of the stellar density). 

Another important consequence of this work is that we find that the analytic
model for the photo-eccentric effect is only valid under the condition that:

\begin{align}
\Bigg(\frac{P}{\mathrm{days}}\Bigg)^{4/3} \gg 0.101 \Bigg( \frac{\rho_{\star,\tru}}{\mathrm{g\,cm}^{-3}} \Bigg)^{-2/3} \Bigg( \frac{1+e}{1-e} \Bigg)^3.
\end{align}

The above has a very steep dependency on $e$ and rapidly rises as $e$ approaches
unity, due to the $(1-e)^{-3}$ term.  This is particularly salient in light of 
the prediction and subsequent observational search for proto-hot Jupiters on 
super-eccentric orbits by \citet{socrates:2012} and \citet{dawson:2013} 
respectively. For example, if we wish to exploit the analytic photo-eccentric 
effect to search for objects with $e=0.9$ around a Solar-like star then we 
require $P\gg114$\,days i.e. we need $P\gtrsim1000$\,d. If this condition is not 
satisfied, the highly eccentric planet would still induce a large 
$(\rho_{\star,\obs}/\rho_{\star,\tru})$ discrepancy, but one cannot reliably use 
the photo-eccentric equations to back out $e$ or $e_{\mathrm{min}}$.

For individual systems, priors on these other AP effects are likely less
relevant since for many well-characterized transiting planet systems there are
often additional observational constraints on many of the terms which
affect the various AP effects e.g. rotational modulations, transit timing
variations, adaptive optics, centroid offsets, etc. Therefore, SAP still
presents arguably the most feasible technique for measuring the eccentricity
of small, habitable-zone planets with current techniques (e.g. see the recent
demonstration with the habitable-zone planet Kepler-22b, 
\citealt{kepler22:2013}).

\subsection{Blend analyses with MAP}
\label{sub:MAP}

Throughout this work we have derived analytic results for various AP effects
in the Single-body Asterodensity Profiling (SAP) paradigm, since all results 
compare the observed stellar density to some independent ``true'' measure.
However, it is trivial to extend our results to the case of Multi-body 
Asterodensity Profiling (MAP), which was first discussed in \citet{map:2012}.
By comparing the observed stellar density between transiting planets $j$
and $k$, and assuming the objects orbit the same star, one is able to
extract information on the state of a system. Since the issue of eccentricity
is discussed in detail in \citet{map:2012}, we do not repeat the arguments
made in that work but briefly summarize that the authors found an analytic
minimum constraint on the pair-wise sum of eccentricities for planets $j$ and
$k$ is easily derived using MAP:

\begin{align}
e_j + e_k \geq \frac{ \Theta_{jk} - 1}{2},
\end{align}

where

\begin{align}
\Theta_{jk} &\equiv \Bigg( \frac{ \rho_{\star,\obs,j} }{ \rho_{\star,\obs,k} } \Bigg)^{2/3}.
\end{align}

The two-thirds index was chosen since it naturally removes a three-halves index
in the expression for the photo-eccentric effect. As shown in this work, it
can be seen that the photo-blend effect also happens to be described by a 
three-halves power and thus the same $\Theta_{jk}$ definition is valuable
in performing MAP for blend analysis. This is pertinent since the photo-blend
effect can induce very large deviations in the observed stellar densities and,
for heavily blended systems, this effect dominates AP. Using 
Equation~\ref{eqn:photoblend} then, one may write the MAP blend equation as:

\begin{align}
\Theta_{jk} &= \frac{ [ (1+\sqrt{\mathcal{B}} p_{\obs,j})^2 - b_{\obs,j}^2 ] [ (1+p_{\obs,k})^2-b_{\obs,k}^2]}{ [ (1+\sqrt{\mathcal{B}} p_{\obs,k})^2 - b_{\obs,k}^2 ] [ (1+p_{\obs,j})^2-b_{\obs,j}^2]}.
\end{align}

It is possible to invert this equation and actually solving for 
$\sqrt{\mathcal{B}}$, yielding two roots from a quadratic equation:

\begin{align}
\sqrt{\mathcal{B}_{\pm}^{\mathrm{MAP}}} &= \frac{ p_{\obs,j} \alpha_{\obs,k} - \Theta_{jk} p_{\obs,k} \alpha_{\obs,j} \pm \sqrt{\mathcal{D}_{jk}} }{ \Theta_{jk} p_{\obs,k}^2 \alpha_{\obs,j} - p_{\obs,j}^2 \alpha_{\obs,k}},
\end{align}

where we make the substitutions

\begin{align}
\alpha_k &= (1+p_{\obs,k})^2 - b_{\obs,k}^2 ,\\
\alpha_j &= (1+p_{\obs,j})^2 - b_{\obs,j}^2 ,\\
\mathcal{D}_{jk} &= b_{\obs,k}^2 p_{\obs,k}^2 \alpha_{\obs,j}^2 \Theta_{jk}^2 - b_{\obs,j}^2 p_{\obs,j}^2 \alpha_{\obs,k}^2 \nonumber\\
\qquad& - \alpha_{\obs,k} \alpha_{\obs,j} \Theta_{jk} \Big[ 2 p_{\obs,k} p_{\obs,j} \nonumber\\
\qquad& - p_{\obs,k}^2 (1-b_{\obs,j}^2) - p_{\obs,j}^2 (1-b_{\obs,k}^2) \Big].
\end{align}

As useful check of these expressions is to evaluate them in the limit of 
$\Theta_{jk}\rightarrow1$, which would be the observed value in the absence of
any blending:

\begin{align}
&\lim_{\Theta_{jk} \rightarrow 1} \sqrt{\mathcal{B}_{-}^{\mathrm{MAP}}} = 1 ,\\
&\lim_{\Theta_{jk} \rightarrow 1} \sqrt{\mathcal{B}_{+}^{\mathrm{MAP}}} = \Big[ p_{\obs,k} (2+p_{\obs,k}) (1-b_{\obs,j}^2) \nonumber\\
\qquad& - p_{\obs,j} (2+p_{\obs}) (1-b_{\obs,k}^2) \Big] \Big[ p_{\obs,j}^2 (1-b_{\obs,k}^2) \nonumber\\
\qquad& - p_{\obs,k}^2 (1-b_{\obs,j}^2) + 2 p_{\obs,k} p_{\obs,j} (p_{\obs,j}-p_{\obs,k})\Big]^{-1}.
\end{align}

Therefore, as expected, we recover the $\mathcal{B}=1$ solution corresponding to
no blending. However, the second root cannot be trivially dismissed and is
physically plausible. Generating uniform random values for $0<p_{\obs,j}<1$,
$0<p_{\obs,k}<1$, $0<b_{\obs,j}<(1-p_{\obs,j})$ and 
$0<b_{\obs,k}<(1-p_{\obs,k})$, we find that 68.3\% of the samples lie in
the range $1.4<\mathcal{B}_{+}^{\mathrm{MAP}}<14.4$ with a median of 3.6.
Therefore, in the limit of $\Theta_{jk}\rightarrow1$, the 
$\mathcal{B}_{+}^{\mathrm{MAP}}$ solution does not produce grossly large
blend factors which can be easily dismissed as unphysical.

We therefore conclude that for 2-planet systems, blend analyses with MAP
will be challenged by this apparent bi-modality. However, with $n\geq3$
planets, one may derive $n!/(2!(n-2)!)$ pair-wise combinations of $\Theta_{jk}$
and the true solution for $\mathcal{B}$ will be recovered in one of the two
roots every time. Therefore, it should be possible to identify which root
corresponds to the true solution in 3 or more planet solutions by root
comparison. We leave more detailed investigations of MAP blend analysis to
future studies.

\subsection{Future Work}

We hope that the investigations presented in this paper provide the 
foundational analytic theory for asterodensity profiling, but we are acutely 
aware that there is a great deal of theoretical and observational work still to 
accomplish in this new area of study. To begin with, there are numerous ignored
effects known to distort the transit light curve and thus have the potential
to impart AP signatures, such as planetary rings \citep{ohta:2009,barnes:2004}, 
planetary oblateness \citep{carter:2010}, atmospheric lensing \citep{hui:2002},
etc. We also did not consider cases where $p>1$, such as total eclipses of
white dwarfs discussed in \citet{agol:2011}.

In order to retrieve the eccentricity distribution using AP, we suggest
that significant work is needed to understand the blend distribution, TTV
distribution, starspot distribution, etc in order to adequately deconvolve
the contribution from other AP effects. Well-characterized individual systems
will likely be less dependent upon these prior distributions and so immediate
observational progress can surely be made here (e.g. 
\citealt{dawson:2012b,kepler22:2013}). In such cases, we would advocate
research into how well the blend factor can be constrained and whether systems
can be practically validated using AP.

Despite these challenges, we envisage that AP can be a powerful tool for 
archival \emph{Kepler} data and for the forthcoming TESS mission, for both
measuring the eccentricity distribution and validating/vetting planetary
candidates. This work also underscores the valuable symbiosis between exoplanet
transits and asteroseismology for characterizing distant worlds.

\section*{Acknowledgments}

DMK would like to thank D. Sliski, S. Ballard and J. Irwin for very helpful
conversations in the preparation of this work. Thanks to the anonymous reviewer
for their constructive comments. DMK is supported by the NASA Sagan Fellowship.

%%%%%%%%%%%%%%%%%%%%%%%%%%%%%%%%%%%%%%%%%%%%%%%%%%%%%%%%%%%%%%%%%%%%%%%%%%%%%%%%
%%%%%%%%%%%%%%%%%%%%%%%%%%%%%%%%%%%%%%%%%%%%%%%%%%%%%%%%%%%%%%%%%%%%%%%%%%%%%%%%
%%%%%%%%%%%%%%%%%%%%%%%%%%%%%%%%%%%%%%%%%%%%%%%%%%%%%%%%%%%%%%%%%%%%%%%%%%%%%%%%
%%%%%%%%%%%%%%%%%%%%%%%%%%%%%%%%%%%%%%%%%%%%%%%%%%%%%%%%%%%%%%%%%%%%%%%%%%%%%%%%
%%%%%%%%%%%%%%%%%%%%%%%%%%%%%%%%%%%%%%%%%%%%%%%%%%%%%%%%%%%%%%%%%%%%%%%%%%%%%%%%
%%%%%%%%%%%%%%%%%%%%%%%%%%%%%%%%%%%%%%%%%%%%%%%%%%%%%%%%%%%%%%%%%%%%%%%%%%%%%%%%
%%%%%%%%%%%%%%%%%%%%%%%%%%%%%%%%%%%%%%%%%%%%%%%%%%%%%%%%%%%%%%%%%%%%%%%%%%%%%%%%
%%%%%%%%%%%%%%%%%%%%%%%%%%%%%%%%%%%%%%%%%%%%%%%%%%%%%%%%%%%%%%%%%%%%%%%%%%%%%%%%
%%%%%%%%%%%%%%%%%%%%%%%%%%%%%%%%%%%%%%%%%%%%%%%%%%%%%%%%%%%%%%%%%%%%%%%%%%%%%%%%

\clearpage
\appendix

\section{Derivation of the Photo-blend Effect}
\label{app:photoblend}

\subsection{Ratio-of-radii bias}

We here provide a formal derivation for the photo-blend effect, followed by
in subsequent appendices by other relevant derivations of important results
presented in this paper. All derivations, unless otherwise stated, will
follow the methodology outlined in \S\ref{sub:method}. Further, for each
derivation of a specific AP effect, all other AP effects will be ignored in
order to provide results for each effect in isolation.

We begin by considering blends. By virtue of the definition of the blend factor
given in Equation~\ref{eqn:Beqn}, the transit depth of a blended source will be 
diminished by the factor $\mathcal{B}$. It therefore follows that the derived 
ratio-of-radii, $p$, is affected via

\begin{align}
p_{\obs} = p_{\tru}/\sqrt{\mathcal{B}},
\label{eqn:pobs}
\end{align}

where we use the subscripts ``true'' and ``obs'' to distinguish between the 
truth and that which one naively adopts the standard simple assumptions of no 
blend, a circular orbit, etc. The true value may be therefore be retrieved using

\begin{align}
p_{\tru} = p_{\obs} \sqrt{\mathcal{B}}.
\label{eqn:ptrue}
\end{align}

\subsection{Impact parameter bias}
\label{sub:bobs}

As stated earlier, we will ignore all other AP effects in what follows. 
Accordingly, one may define the transit impact parameter as a function of just 
three terms $T_{14}$, $T_{23}$ and $p$, as demonstrated by \citet{seager:2003}. 
In this framework, $T_{14}$ is the first-to-fourth contact duration and $T_{23}$ 
is the second-to-third contact duration. Critically, these durations are 
unaffected by the act of a blend \citep{kippingtinetti:2010}. The same statement 
can also be said of the the orbital period which is calculated by the interval 
between transits:

\begin{align}
T_{14,\obs} &= T_{14,\tru} = T_{14},\nonumber\\
T_{23,\obs} &= T_{23,\tru} = T_{23},\nonumber\\
P_{\obs} &= P_{\tru} = P,
\end{align}

where we drop the explicit ``true'' subscript on the right-hand side (RHS). We 
follow this pattern in what follows, where the reader should interpret any term 
missing an explicit true/obs subscript to imply that we are referring to the 
true value. Having now defined the effect of blends on each of the key 
observable terms, we may now feed our expressions for $p_{\obs}$, $T_{14,\obs}$ 
and $T_{23,\obs}$ into Equation~\ref{eqn:bseager} from \citet{seager:2003} to 
derive the observed impact parameter:

\begin{align}
b_{\obs}^2 &= \frac{ (1-p_{\tru} \mathcal{B}^{-1/2})^2 - \frac{\sin^2(T_{23}\pi/P)}{\sin^2(T_{14}\pi/P)} (1+p_{\tru} \mathcal{B}^{-1/2})^2 }{ 1 - \frac{\sin^2(T_{23}\pi/P)}{\sin^2(T_{14}\pi/P)} }.
\label{eqn:bobsblend1}
\end{align}

Let us assume that $\mathcal{B}=1$ i.e. no blend is present. Plugging equations 
Equation~\ref{eqn:durations} into Equation~\ref{eqn:bobsblend1} in this limit
yields

\begin{align}
\lim_{B \to 1} b_{\obs}^2 &= b_{\tru}^2,
\end{align}

as expected. Now consider that a blend source is present. Again feeding
Equation~\ref{eqn:durations} Equation~\ref{eqn:bobsblend1} yields (without
any approximation):

\begin{align}
b_{\obs}^2 &= \frac{ \mathcal{B} + p_{\tru}^2 - \sqrt{\mathcal{B}} ( 1 + p_{\tru}^2 - b_{\tru}^2 ) }{ \mathcal{B} }.
\label{eqn:bobssy}
\end{align}

Since the above expression clearly scales with $b_{\tru}$, then we may find
the maximum/minimum range of the above by evaluating when $b_{\tru}\rightarrow
b_{\tru,\mathrm{min}}=0$ and $b_{\tru}\rightarrow
b_{\tru,\mathrm{max}}=1-p_{\tru}$:

\begin{align}
\frac{ \mathcal{B}+p_{\tru}^2 }{ \mathcal{B} } - \frac{ 1+p_{\tru}^2 }{ \sqrt{\mathcal{B}} } \leq b_{\obs}^2 \leq \frac{ \mathcal{B}+p_{\tru}^2 }{ \mathcal{B} } - \frac{ 2p_{\tru} }{ \sqrt{\mathcal{B}} }
\label{eqn:bobslimits}
\end{align}

If we replace $p_{\tru}$ with $p_{\obs} \sqrt{B}$ then the RHS simplifies to
$b_{\obs}\leq (1-p-{\obs})$, displaying an analogous form the the boundary
condition imparted on the true impact parameter.
In the limit of no blending and extreme blending, Equation~\ref{eqn:bobslimits}
gives:

\begin{align}
0 \leq &\lim_{\mathcal{B}\to1} b_{\obs}^2 \leq (1-p_{\tru})^2 \nonumber\\
1 \leq &\lim_{\mathcal{B}\to\infty} b_{\obs}^2 \leq 1.
\end{align}

Therefore, we find that $0\leq b_{\obs} \leq (1-p_{\tru})$. Recall that we
have also showed that $0\leq b_{\obs} \leq (1-p_{\obs})$. Since both statements
are true, one must take precedent over the other and since $p_{\obs}<p_{\tru}$
then the latter limit is the more constraining one i.e. 
$0\leq b_{\obs} \leq (1-p_{\obs})$ for all $\mathcal{B}\geq$ and 
$0\leq b_{\tru}\leq(1-p_{\tru})$.

Equation~\ref{eqn:bobssy} may be re-written by replacing the $p_{\tru}$ terms 
with the observed values to give

\begin{align}
b_{\obs}^2 &= 1 + p_{\obs}^2 - \Bigg( \frac{ 1 + \mathcal{B} p_{\obs}^2 - b_{\tru}^2 }{ \sqrt{\mathcal{B}} } \Bigg).
\label{eqn:bobs}
\end{align}

The inverse of this expression is easily shown to be:

\begin{align}
b_{\tru}^2 &= 1 + \mathcal{B} p_{\obs}^2 - \sqrt{\mathcal{B}} (1-b_{\obs}^2) - p_{\obs}^2.
\end{align}

\subsection{Scaled semi-major axis bias}

The scaled semi-major axis, $(a/R_{\star})$, can also be derived from the
observed durations and ratio-of-radii via Equation~\ref{eqn:aRseager}, to give:

\begin{align}
(a/R_{\star})_{\obs}^2 &= \frac{(1 + p_{\obs})^2 - b_{\obs}^2 (1-\sin^2(T_{14}\pi/P)}{ \sin^2(T_{14}\pi/P) }.
\end{align}

In the limit of no blend, then $b_{\obs} \rightarrow b_{\tru}$ and 
$p_{\obs} \rightarrow p_{\tru}$, giving the expected result that

\begin{align}
\lim_{\mathcal{B} \to 1} (a/R_{\star})_{\obs}^2 &= (a/R_{\star})_{\tru}^2.
\end{align}

However, in the case of a non-unity blend factor, we find

\begin{align}
&(a/R_{\star})_{\obs}^2 = \Bigg(\frac{(a/R_{\star})^2-b^2}{(1+p)^2-b^2}\Bigg) \Bigg( (1+p)^2 \nonumber\\
\qquad& - \frac{ [ (a/R_{\star})^2 - (1+p)^2 ][ B + p^2 - \sqrt{B} (1+p^2-b^2) ] }{ B [(a/R_{\star})^2 - b^2 ] } \Bigg).
\label{eqn:aRobs}
\end{align}

\subsection{Mean stellar density bias}

Finally, we come to the parameter of interest, the mean stellar density, 
$\rho_{\star,\obs}$. Following the definition in Equation~\ref{eqn:rhoseager},
we have:

\begin{align}
\rho_{\star,\obs} &\equiv \frac{ 3\pi (a/R_{\star})_{\obs}^{3/2} }{ G P^2 }.
\end{align}

As before, for an unblended target star we recover

\begin{align}
\lim_{\mathcal{B} \to 1} \rho_{\star,\obs} &= \rho_{\star,\tru}.
\end{align}

For blended planets, the equation is more complicated, particular when we
make the substitution that $(a/R_{\star})_{\tru}^3 = (G P^2 \rho_{\star,\tru})/
(3\pi)$.

By inspection of Equation~\ref{eqn:aRobs}, we found that assuming 
$(a/R_{\star})^2\gg (1+p)^2$ (which also imparts $(a/R_{\star})^2\gg b^2$ since 
$b<(1+p)$ in order for a transit to occur) allows for significant reduction in 
the form of the expression for $(\rho_{\star,\obs}/\rho_{\star,\tru})$ to:

\begin{align}
\lim_{(a/R_{\star})^2\gg(1+p)^2} \Big(\frac{\rho_{\star,\obs}}{\rho_{\star,\tru}}\Big) &= \Bigg(\frac{(1+p)^2-b_{\obs}^2}{(1+p)^2-b^2}\Bigg)^{3/2}
\label{eqn:rhoobs1}
\end{align}

Since we assume that the transit displays a flat bottom, then $b^2\lesssim1$
and $p\lesssim1$ giving us $(1+p)^2\lesssim4$ so that our assumption becomes
$(a/R_{\star})^2\gg4$. By definition, $(a/R_{\star})>1$ at all times in order
for the transit to be physical (otherwise the transiting object is inside the
star). Whilst the majority of exoplanets easily satisfy the condition that
$(a/R_{\star})^2\gg4$, some very short-period objects such as Kepler-78b have
$(a/R_{\star})^2\sim3$ \citep{sanchis:2013}. $(a/R_{\star})$ may be estimated
using $P$ and $\rho_{\star,\tru}$ and converting into typical units of measure
we determine that our approximation is valid for:

\begin{align}
\Big(\frac{P}{\mathrm{days}}\Big)^{4/3} \gg 0.389\,\Big(\frac{\rho_{\star,\tru}}{\mathrm{g\,cm}^3}\Big)^{-2/3}
\end{align}

Under this condition then, one may re-write Equation~\ref{eqn:rhoobs1} in terms
of the observables:

\begin{align}
\lim_{(a/R_{\star})^2\gg(1+p)^2} \Big(\frac{\rho_{\star,\obs}}{\rho_{\star,\tru}}\Big) &= \Bigg(\frac{(1+\sqrt{\mathcal{B}} p_{\obs})^2-b_{\obs}^2}{\sqrt{\mathcal{B}} ( (1+p_{\obs})^2-b_{\obs}^2 )}\Bigg)^{3/2}.
\label{eqn:rhoobs2}
\end{align}

\subsection{Solving for the blend parameter}
\label{sub:blendinversion}

Since the observer directly determines $p_{\obs}$ and $b_{\obs}$, then 
Equation~\ref{eqn:rhoobs2} suggests that one should be able to invert the
above and infer $\mathcal{B}$. However, in doing so, we find that one recovers
a quadratic solution and both roots are ostensibly plausible:

\begin{align}
\mathcal{B}_{+,-} =& \frac{1}{4p_{\obs}^4} \Bigg( -2p_{\obs} + \Big(\frac{\rho_{\star,\obs}}{\rho_{\star,\tru}}\Big)^{2/3} [(1+p_{\obs})^2-b_{\obs}^2] \nonumber\\
\qquad& \pm \Bigg[ \Big( p_{\obs} \Big[(2+p_{\obs})\Big(\frac{\rho_{\star,\obs}}{\rho_{\star,\tru}}\Big)^{2/3}-2\Big] \nonumber\\
\qquad& + \Big(\frac{\rho_{\star,\obs}}{\rho_{\star,\tru}}\Big)^{2/3} (1-b_{\obs}^2) \Big)^2 - 4 p_{\obs}^2 (1-b_{\obs}^2) \Bigg]^{1/2} \Bigg)
\label{eqn:Bplusminusapp}
\end{align}

There are several analytic insights that can be made with this expression.
Firstly, at the extreme solution of $(\rho_{\star,\obs}/\rho_{\star,\tru})=1$, 
example plots of the functions (e.g. see Figure~\ref{fig:Bcontours}) show that 
these points correspond to the maximum and minimum in $\mathcal{B}$-space:

\begin{align}
\mathcal{B} \geq \lim_{(\rho_{\star,\obs}/\rho_{\star,\tru})\rightarrow1} \mathcal{B}_{-} &= 1,\\
\mathcal{B} \leq \lim_{(\rho_{\star,\obs}/\rho_{\star,\tru})\rightarrow1} \mathcal{B}_{+} &= \frac{(1 - b_{\obs}^2)^2}{p_{\obs}^4}.
\end{align}

We also note that two functions, $\mathcal{B}_{+}$ and $\mathcal{B}_{-}$, meet
at an apparent minimum in $(\rho_{\star,\obs}/\rho_{\star,\tru})$-space. This
point is found by solving 
$\partial(\rho_{\star,\obs}/\rho_{\star,\tru})/\partial B=0$ for $B$, giving:

\begin{align}
\lim_{\big(\frac{\rho_{\star,\obs}}{\rho_{\star,\tru}}\big)\rightarrow\big(\frac{\rho_{\star,\obs}}{\rho_{\star,\tru}}\big)_{\mathrm{min}}} \mathcal{B}=\frac{1-b_{\obs}^2}{p_{\obs}^2} ,
\end{align}

which may be used to determine the equivalent location in 
$(\rho_{\star,\obs}/\rho_{\star,\tru})$-space, corresponding to the maximum
and minimum limits of said parameter:

\begin{align}
\Big(\frac{\rho_{\star,\obs}}{\rho_{\star,\tru}}\Big)\geq&\Big(\frac{\rho_{\star,\obs}}{\rho_{\star,\tru}}\Big)_{\mathrm{min}} = \Bigg(
\frac{2 p_{\obs} (1+\sqrt{1-b_{\obs}^2})}{ (1+p_{\obs})^2 - b_{\obs}^2 } \Bigg)^{3/2},\\
\Big(\frac{\rho_{\star,\obs}}{\rho_{\star,\tru}}\Big)\leq&\Big(\frac{\rho_{\star,\obs}}{\rho_{\star,\tru}}\Big)_{\mathrm{max}} = 1.
\label{eqn:rhomaxmin}
\end{align}

Crucially then, a measurement of $(\frac{\rho_{\star,\obs}}{\rho_{\star,\tru}})$
below the minimum or above the maximum should not be possible for any degree of 
blending. Such a case therefore would mean that another AP effect is responsible 
for the deviation (which realistically can only be the photo-eccentric effect) 
or the star hosting the eclipsing body does not possess a true stellar density 
equal to $\rho_{\star,\tru}$.

Finally, we note that the contours never cross the small-planet limit
found by evaluating Equation~\ref{eqn:rhoobs2} in the limit $p_{\obs}\ll1$:

\begin{align}
\lim_{p_{\obs}\ll1} \Bigg(\lim_{(a/R_{\star})^2\gg(1+p)^2} \Big(\frac{\rho_{\star,\obs}}{\rho_{\star,\tru}}\Big)\Bigg) &= \mathcal{B}^{-3/4}.
\label{eqn:rhoobssmall}
\end{align}

\subsection{Allowed Range of $\mathcal{B}_{-}$}

Plotting some example functions in Figure~\ref{fig:Bcontours} reveals that
the $\mathcal{B}_{+}$ solutions extend up to suspiciously high $\mathcal{B}$.
This issue can be phrased mathematically by computing the true value of $p$
once one corrects for the blending factor, given by Equation~\ref{eqn:ptrue}.
Since a fundamental assumption of our work is that a flat-bottomed transit is 
observed, then we expect $p_{\tru}<(1-b_{\tru})$ at all times and since the
minimum value of $b_{\tru}$ is zero then the maximum limit is $p_{\tru}<1$.
By this criteria and inspection of the contours in Figure~\ref{fig:Bcontours},
we note that there appear to be some apparently forbidden $p_{\tru}$ values
along the $\mathcal{B}_{+}$ contour with $\mathcal{B}$ reaching $\sim10^8$.

Before exploring the very high blend factors produced by the $\mathcal{B}_{+}$
root, we first evaluate the maximum possible $p_{\tru}$ value along the 
$\mathcal{B}_{-}$ contour, which occurs at the point where the $\mathcal{B}_{-}$ 
meets $\mathcal{B}_{+}$ i.e. when 
$\big(\frac{\rho_{\star,\obs}}{\rho_{\star,\tru}}\big)\rightarrow 
\big(\frac{\rho_{\star,\obs}}{\rho_{\star,\tru}}\big)_{\mathrm{min}}$:

\begin{align}
\lim_{\big(\frac{\rho_{\star,\obs}}{\rho_{\star,\tru}}\big)\rightarrow\big(\frac{\rho_{\star,\obs}}{\rho_{\star,\tru}}\big)_{\mathrm{min}}} p_{\tru} &= \sqrt{1 - b_{\obs}^2}
\end{align}

We have already derived an expression for $b_{\obs}$ earlier in 
Equation~\ref{eqn:bobs}, which scales with $b_{\tru}$. It was shown earlier
than $0\leq b_{\obs}\leq(1-p_{\obs})$ for the allowed parameter range considered 
in this study. This means that $p_{\tru}<\sqrt{1-(1-p_{\obs})^2}\leq1$ since 
$p_{\obs}\leq1$. This therefore proves that all loci along the $\mathcal{B}_{-}$
contour reside in unforbidden parameter space.

\subsection{Allowed Range of $\mathcal{B}_{+}$}

It is easy to show that at least some of the loci along the $\mathcal{B}_{+}$
contour produce $p_{\tru}>1$ and thus break the fundamental assumptions of
our work. For example, consider the maximum possible of $\mathcal{B}_{+}$ found 
when $(\rho_{\star,\obs}/\rho_{\star,\tru})\rightarrow1$, as mentioned earlier:

\begin{align}
\lim_{(\rho_{\star,\obs}/\rho_{\star,\tru})\rightarrow1} \mathcal{B}_{+} &= \frac{(1 - b_{\obs}^2)^2}{p_{\obs}^4}.
\end{align}

Requiring $p_{\tru}<1$ is equivalent to $\mathcal{B}<p_{\obs}^{-2}$ which means
that in order for the above satisfy this we require 
$(1 - b_{\obs}^2) < p_{\obs}$. However, since $b_{\obs}<1-p_{\obs}$ as shown
earlier in Appendix~\ref{sub:bobs}, then this condition can \emph{never}
be in effect. Therefore, there is no doubt that $\mathcal{B}_{+}$ at least
partially samples forbidden parameter space.

We may actually solve for the point along $\mathcal{B}_{+}$ when this
breakdown occurs. This must occurs when $\mathcal{B}_{+} = p_{\obs}^{-2}$
since we require $p_{\tru}<1$ which implies $\mathcal{B}<p_{\obs}^{-2}$ at all
times. Solving this expression for $(\rho_{\star,\obs}/\rho_{\star,\tru})$ 
yields a quadratic equation with two roots. The first root has the solution:

\begin{align}
\Big(\frac{\rho_{\star,\obs}}{\rho_{\star,\tru}}\Big) &= \Bigg( \frac{ b_{\obs}^2 p_{\obs} }{ (1+p_{\obs})^2-b_{\obs}^2 } \Bigg)^{3/2}.
\end{align}

This may be compared to the minimum allowed value of 
$(\rho_{\star,\obs}/\rho_{\star,\tru})$ derived earlier in 
Equation~\ref{eqn:rhomaxmin}, meaning that we require:

\begin{align}
&\Bigg( \frac{ b_{\obs}^2 p_{\obs} }{ (1+p_{\obs})^2-b_{\obs}^2 } \Bigg)^{3/2} > 
\Bigg( \frac{ 2 p_{\obs} (1+\sqrt{1-b_{\obs}^2}) }{ (1+p_{\obs})^2-b_{\obs}^2 } \Bigg)^{3/2},\nonumber\\
&\Rightarrow b_{\obs}^2 \ngtr 2 (1+\sqrt{1-b_{\obs}^2})\,\,\forall\,\,0<b_{\obs}<1
\end{align}

where the second line shows the condition is not satisfied (in fact the exact
opposite condition is in effect). This allows us to summarily reject this root
as a genuine solution. The remaining root has the form:

\begin{align}
\Big(\frac{\rho_{\star,\obs}}{\rho_{\star,\tru}}\Big)_{\mathcal{B}_{+,\mathrm{max}}} &= \Bigg( \frac{ (4-b_{\obs}^2) p_{\obs} }{ (1+p_{\obs})^2-b_{\obs}^2 } \Bigg)^{3/2},
\end{align}

which does satisfy the condition of being greater than the minimum estimate in
Equation~\ref{eqn:rhomaxmin} for all $b_{\obs}>0$. This maximum limit is marked
with gray circles on the example plots shown in Figure~\ref{fig:Bcontours}.
Therefore, $\mathcal{B}_{+}$ produces is a valid solution when we have:

\begin{align}
\Big(\frac{\rho_{\star,\obs}}{\rho_{\star,\tru}}\Big)_{\mathrm{min}} < \Big(\frac{\rho_{\star,\obs}}{\rho_{\star,\tru}}\Big) < \Big(\frac{\rho_{\star,\obs}}{\rho_{\star,\tru}}\Big)_{\mathcal{B}_{+,\mathrm{crit}}}.
\end{align}

A measurement of the density in this range means that the inverse solution for
$\mathcal{B}$ has two roots. Therefore, one should expect a bi-modal posterior
distribution for $\mathcal{B}$ when using SAP in such cases, provided the prior 
range in $\mathcal{B}$ is allowed to explore to high blend factors. We also note
that in the the limit of $b_{\obs}^2\rightarrow0$ then 
$\big(\frac{\rho_{\star,\obs}}{\rho_{\star,\tru}}\big)_{\mathrm{min}}
=\big(\frac{\rho_{\star,\obs}}{\rho_{\star,\tru}}\big)_{\mathcal{B}_{+,
\mathrm{crit}}}$ meaning the $\mathcal{B}_{+}$ solution is always forbidden in
such a case. This is also evident in the top-left panel of 
Figure~\ref{fig:Bcontours} where the two curves corresponding to these limits
overlap.

\section{Valid Range for the Analytic Photo-eccentric Effect}
\label{app:psi}

\subsection{The Transit Duration Equation}

In this paper, we assume that the observed stellar density is affected by
the photo-eccentric effect via a simple analytic formula. In this section,
we investigate under what conditions this simple formula is actually a valid
since there appears to be no previous efforts to quantify the validity of this 
crucial assumption. The observed stellar density is assumed to behave as
\citep{investigations:2010}:

\begin{align}
\Big(\frac{\rho_{\star,\obs}}{\rho_{\star,\tru}}\Big) &= \Psi,
\label{eqn:photoe1}
\end{align}

where we define

\begin{align}
\Psi \equiv \frac{(1-e\sin\omega)^3}{(1-e^2)^{3/2}}.
\end{align}

Inferences about the eccentricity of a system made using the above expressions
are defined here as exploiting the analytic photo-eccentric.
These expressions are taken from \citet{investigations:2010} but we note that
many other authors have used this function for modeling the 
photo-eccentric effect \citep{winn:2010,carter:2011,map:2012,dawson:2012a}. 
Given the widespread use of this expression, it is crucial to understand
the limits of the equation in question. The expressions above
are derived by setting $T_{14,\obs}$ and $T_{23,\obs}$ to that expected for
a planet with orbital eccentricity, $e$, and argument of periastron, $\omega$.
To date, there is no known exact analytic expression for the duration of
a transit on an eccentric orbit but \citet{investigations:2010} derived an
approximate expression, provided by Equation~15 of that work:

\begin{align}
T_{ _{23}^{14} } &= \frac{P}{\pi} \frac{\varrho_c^2}{\sqrt{1-e^2}} \sin^{-1} 
\sqrt{ \frac{(1\pm p)^2-b^2}{ (a/R_{\star})^2 \varrho_c^2 - b^2 } },
\label{eqn:Teccentric}
\end{align}

where \citet{investigations:2010} define

\begin{align}
\varrho_c &\equiv \frac{1-e^2}{1+e\sin\omega}.
\end{align}

\citet{investigations:2010} demonstrate that Equation~\ref{eqn:Teccentric}
is an excellent approximation to the true transit duration (which can be 
computed more laboriously via the method described in \citealt{kipping:2008}).
As demonstrated in \citet{investigations:2010}, these approximate expressions
become most erroneous when $(a/R_{\star})$ is small. However, even at
$(a/R_{\star})=5$, the expression performs better than 1\% accuracy across the
vast majority of parameter space and the paper finds an impressive average
accuracy of $<0.1$\% for $|e\sin\omega|<0.5$ and $|e\cos\omega|<0.85$.
Compared to the other assumptions made in deriving Equation~\ref{eqn:photoe1},
which we will shortly discuss, Equation~\ref{eqn:Teccentric} is unlikely to
ever be the bottleneck in accuracy. 

Although \citet{investigations:2010} spent great effort exploring the accuracy
of Equation~\ref{eqn:Teccentric}, no effort is spent on the accuracy of the
most relevant equation for the analytic photo-eccentric effect i.e.
Equation~\ref{eqn:photoe1}. The reason for this is quite simply that the
photo-eccentric effect had not been envisaged at this time and so the importance
of Equation~\ref{eqn:photoe1} was not realized. Therefore, we devote this 
section to addressing this important question.

\subsection{Accuracy of the Impact Parameter Approximation}

As with other asterodensity effects, not only is $(a/R_{\star})_{\obs}$ (and 
thus $\rho_{\star,\obs}$) deviant from the truth, but also the observed impact
parameter, $b_{\obs}$, is deviant. Using Equation~\ref{eqn:Teccentric} and
the original \citet{seager:2003} equations, \citet{investigations:2010} (see 
Equation~33) showed that (without any approximations):

\begin{align}
&b_{\obs}^2 = 1 + p^2 + 2p \nonumber \\
&\Bigg(\frac{ \sin^2[\frac{\varrho_c^2}{\sqrt{1-e^2}} \sin^{-1}(\frac{\sqrt{(1-p)^2 - b^2}}{a_R \varrho_c \sin i})] + \sin^2[\frac{\varrho_c^2}{\sqrt{1-e^2}} \sin^{-1}(\frac{\sqrt{(1+p)^2 - b^2}}{a_R \varrho_c \sin i})] }{ \sin^2[\frac{\varrho_c^2}{\sqrt{1-e^2}} \sin^{-1}(\frac{\sqrt{(1-p)^2 - b^2}}{a_R \varrho_c \sin i})] - \sin^2[\frac{\varrho_c^2}{\sqrt{1-e^2}} \sin^{-1}(\frac{\sqrt{(1+p)^2 - b^2}}{a_R \varrho_c \sin i})] }\Bigg).
\label{eqn:bobs_ecc}
\end{align}

\citet{investigations:2010} briefly remark that making a small-angle 
approximation in the trigonometric functions allows one to simplify the above
to $b_{\obs}^2 = b^2$. However, what remains unclear is exactly under what
conditions is a small-angle approximation valid? The small-angle approximation
is actually implemented four times in total, two of which correspond to
$\sin^{-1}x\simeq x$ and two of which correspond to $\sin x \simeq x$. Let us
begin by inspecting the validity of the inverse sine approximation. 

\subsubsection{Accuracy of the inverse sine small-angle approximation}

The inverse sine approximation in question is fully expressed as:

\begin{align}
\sin^{-1}\sqrt{ \frac{ (1\pm p)^2 - b^2 }{ (a/R_{\star})^2 \varrho_c^2 - b^2 } } \simeq \sqrt{ \frac{ (1\pm p)^2 - b^2 }{ (a/R_{\star})^2 \varrho_c^2 - b^2 } }.
\end{align}

It is trivial to show that the $\varrho_c$ term has two extrema at 
$\omega=\pi/2$ and $\omega=3\pi/2$ and so we may consider four distinct
cases under which we require the above approximation to remain true:

\begin{itemize}
\item[{\textbf{[1]}}] $(1\pm p) \rightarrow (1+p)$ and 
$\varrho_c \rightarrow \lim_{\omega\to\pi/2} \varrho_c$ \\
\item[{\textbf{[2]}}] $(1\pm p) \rightarrow (1+p)$ and 
$\varrho_c \rightarrow \lim_{\omega\to3\pi/2} \varrho_c$ \\
\item[{\textbf{[3]}}] $(1\pm p) \rightarrow (1-p)$ and 
$\varrho_c \rightarrow \lim_{\omega\to\pi/2} \varrho_c$ \\
\item[{\textbf{[4]}}] $(1\pm p) \rightarrow (1-p)$ and 
$\varrho_c \rightarrow \lim_{\omega\to3\pi/2} \varrho_c$
\end{itemize}

\begin{figure*}
\begin{center}
\includegraphics[width=16.8 cm]{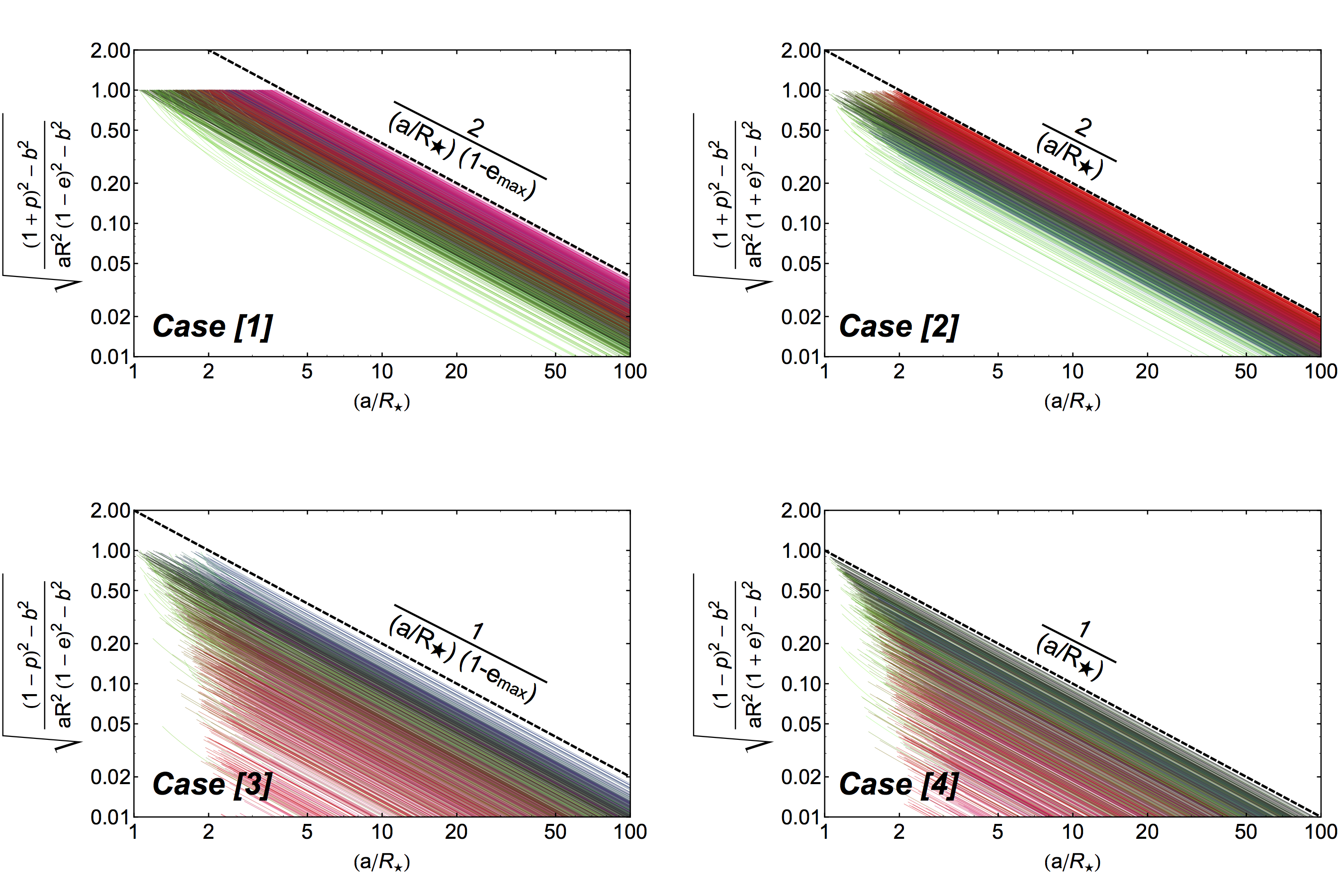}
\caption{\emph{ \textbf{Small-angle inverse sine approximation investigation:}
On the $y$-axis we plot the four extreme possible arguments to the inverse sine
functions present in Equation~\ref{eqn:bobs_ecc}, with respect to 
$(a/R_{\star})$ on the $x$-axis. Each panel shows 1000 random realizations for
$p$, $b$ and $e$, where the RGB-colouring is given by 
$\{\mathrm{R},\mathrm{G},\mathrm{B}\} = \{p,b,e\}$. For each panel, we show
the maximum allowed value of the function in black-dashed. Simulations produced
using $e_{\mathrm{max}}=0.5$, but the upper limits are valid for all 
$0\leq e_{\mathrm{max}}<1$.
}} 
\label{fig:arcsin_tests}
\end{center}
\end{figure*}

In each case, the remaining variables are $(a/R_{\star})$, $b$, $e$ and $p$.
Let us proceed by finding the maximum of the inverse sine function's argument in
all four cases. We demonstrate this by Monte Carlo experiment where we draw
random uniform variates for $0<p<1$, $0<b<(1-p)$ and $0<e<e_{\mathrm{max}}$.
For each realization, we plot the inverse sine argument as a function of the
only remaining dependent variable, $(a/R_{\star})$. We make 1000 plots for each
of the four cases and in each case we determine the maximum value of the
inverse sine function's argument, with respect to $(a/R_{\star})$ and 
$e_{\mathrm{max}}$. In practice this is done by both varying the experiments
for several different $e_{\mathrm{max}}$ values and taking the derivatives of
the inverse sine function's argument. In all Monte Carlo experiments, we enforce
the condition that $(a/R_{\star}) > (1-e)$ to avoid the planet colliding into
the star. Figure~\ref{fig:arcsin_tests} displays our results when we 
arbitarily choose $e_{\mathrm{max}}=0.5$.

After conducting this analysis, for which some results are illustrated in
Figure~\ref{fig:arcsin_tests}, we are able to derive functional upper limits
on the the inverse sine function's argument with respect to $(a/R_{\star})$
and $e_{\mathrm{max}}$. From these four maxima functions, one may use the 
maximum of \emph{these} to demonstrate that:

\begin{align}
&\sqrt{ \frac{ (1\pm p)^2 - b^2 }{ (a/R_{\star})^2 \varrho_c^2 - b^2 } } \leq \frac{2}{(a/R_{\star}) (1-e)}\nonumber\\
\forall\,\,(0\leq p<1);\,\,&(0\leq b<1-p);\,\,(0\leq\omega<2\pi);\,\,(0\leq e<1).
\label{eqn:arcsinmax}
\end{align}

Armed with the above, one may now answer the question as to what range of
orbits the small-angle inverse sine approximation is valid. The Maclaurin
series expansion of the inverse sine function may be expressed as:

\begin{align}
\sin^{-1} x &= x + \frac{x^3}{6} + \mathcal{O}[x^5].
\end{align}

Therefore, the approximation that $\sin^{-1} x \simeq x$ is valid when
$(x^3/6)\ll x$ i.e. when $(x^2/6)\ll 1$. Using our maximum expression for the
inverse sine argument in Equation~\ref{eqn:arcsinmax}, the small-angle
approximation is now valid for:

\begin{align}
&\mathbf{Condition\,\,A} \nonumber \\
&(a/R_{\star})^2 \gg \frac{2}{3} \frac{1}{(1-e)^2}.
\label{eqn:conditionA}
\end{align}

\subsubsection{Accuracy of the sine small-angle approximation}

Let us assume that Condition A is valid so that:

\begin{align}
&\sin \Bigg(\frac{\varrho_c^2}{\sqrt{1-e^2}} \sin^{-1} \sqrt{ \frac{ (1\pm p)^2 - b^2 }{ (a/R_{\star})^2 \varrho_c^2 - b^2 } } \Bigg) \nonumber\\
\qquad& \simeq \sin\Bigg(\frac{\varrho_c^2}{\sqrt{1-e^2}} \sqrt{ \frac{ (1\pm p)^2 - b^2 }{ (a/R_{\star})^2 \varrho_c^2 - b^2 } } \Bigg).
\end{align}

Next, we need to investigate when the small-angle approximation for the sine 
function is valid i.e. when

\begin{align}
&\sin\Bigg(\frac{\varrho_c^2}{\sqrt{1-e^2}} \sqrt{ \frac{ (1\pm p)^2 - b^2 }{ (a/R_{\star})^2 \varrho_c^2 - b^2 } } \Bigg) \nonumber \\
\qquad& \simeq \Bigg(\frac{\varrho_c^2}{\sqrt{1-e^2}} \sqrt{ \frac{ (1\pm p)^2 - b^2 }{ (a/R_{\star})^2 \varrho_c^2 - b^2 } } \Bigg).
\end{align}

As with the investigation of the inverse sine function, we will consider the
four extreme cases of:

\begin{itemize}
\item[{\textbf{[1]}}] $(1\pm p) \rightarrow (1+p)$ and 
$\varrho_c \rightarrow \lim_{\omega\to\pi/2} \varrho_c$ \\
\item[{\textbf{[2]}}] $(1\pm p) \rightarrow (1+p)$ and 
$\varrho_c \rightarrow \lim_{\omega\to3\pi/2} \varrho_c$ \\
\item[{\textbf{[3]}}] $(1\pm p) \rightarrow (1-p)$ and 
$\varrho_c \rightarrow \lim_{\omega\to\pi/2} \varrho_c$ \\
\item[{\textbf{[4]}}] $(1\pm p) \rightarrow (1-p)$ and 
$\varrho_c \rightarrow \lim_{\omega\to3\pi/2} \varrho_c$
\end{itemize}

\begin{figure*}
\begin{center}
\includegraphics[width=16.8 cm]{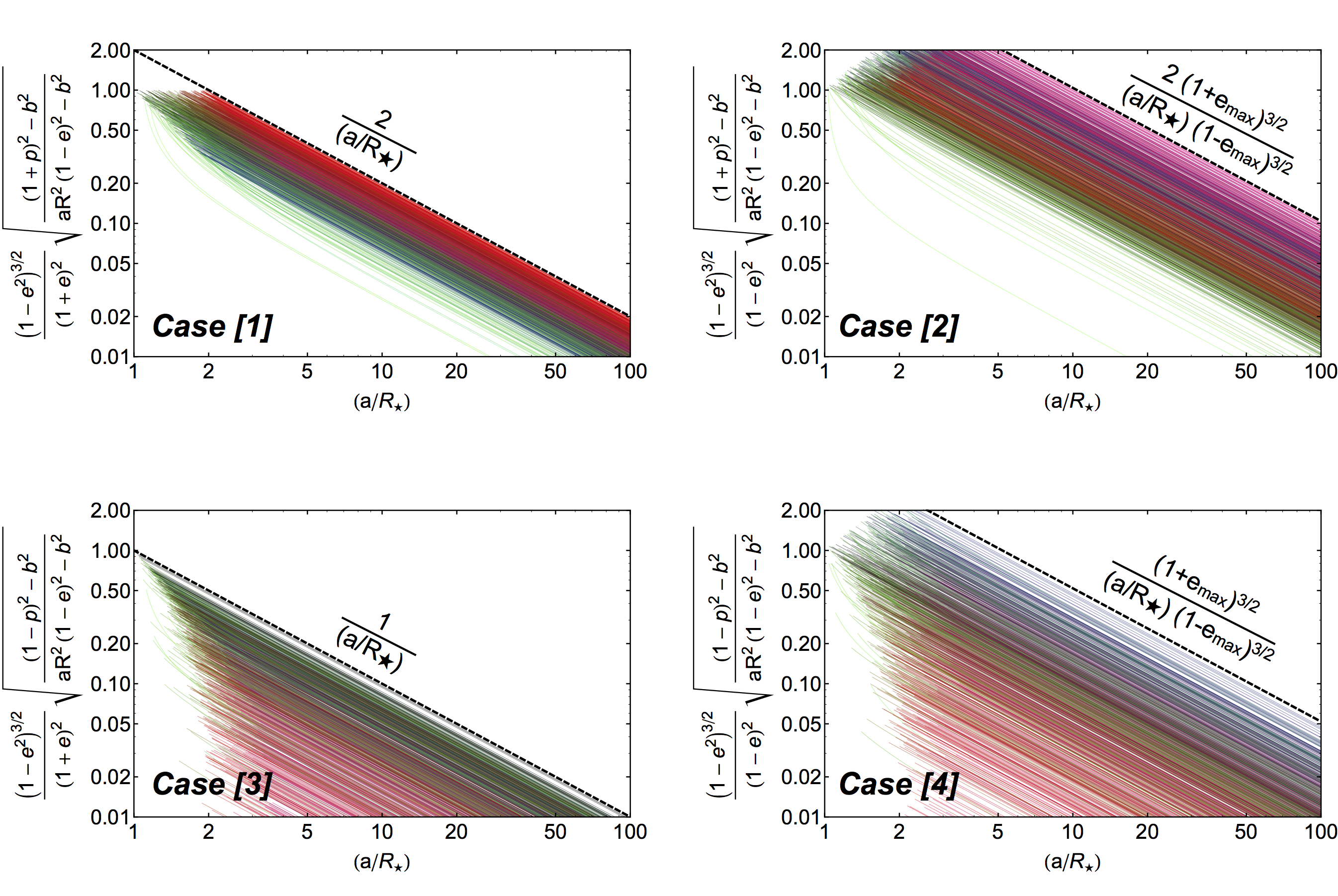}
\caption{\emph{ \textbf{Small-angle sine approximation investigation:}
On the $y$-axis we plot the four extreme possible arguments to the sine
functions present in Equation~\ref{eqn:bobs_ecc}, with respect to 
$(a/R_{\star})$ on the $x$-axis. Each panel shows 1000 random realizations for
$p$, $b$ and $e$, where the RGB-colouring is given by 
$\{\mathrm{R},\mathrm{G},\mathrm{B}\} = \{p,b,e\}$. For each panel, we show
the maximum allowed value of the function in black-dashed. Simulations produced
using $e_{\mathrm{max}}=0.5$, but the upper limits are valid for all 
$0\leq e_{\mathrm{max}}<1$.
}} 
\label{fig:sin_tests}
\end{center}
\end{figure*}

As before, we seek to determine the maximum of the sine function's argument
with respect to $(a/R_{\star})$ and $e_{\mathrm{max}}$ by Monte Carlo 
experiments and analysis of the differentials. Generating random $p$, $b$ and
$e$ values via the same method used earlier, we determine upper limits for
each of the four cases, shown in Figure~\ref{fig:sin_tests}.

After conducting this analysis, we are able to derive functional upper limits
on the the sine function's argument with respect to $(a/R_{\star})$
and $e_{\mathrm{max}}$. From these four maxima functions, one may use the 
maximum of \emph{these} to demonstrate that:

\begin{align}
&\frac{\varrho_c^2}{\sqrt{1-e^2}} \sqrt{ \frac{ (1\pm p)^2 - b^2 }{ (a/R_{\star})^2 \varrho_c^2 - b^2 } } \leq \frac{2}{(a/R_{\star})} \Big(\frac{1+e}{1-e}\Big)^{3/2} \nonumber\\
\forall\,\,(0\leq &p<1);\,\,(0\leq b<1-p);\,\,(0\leq\omega<2\pi);\,\,(0\leq e<1).
\label{eqn:sinmax}
\end{align}

Armed with the above, one may now answer the question as to what range of
orbits the small-angle sine approximation is valid. The Maclaurin
series expansion of the inverse sine function may be expressed as:

\begin{align}
\sin x &= x - \frac{x^3}{6} + \mathcal{O}[x^5].
\end{align}

Therefore, the approximation that $\sin x \simeq x$ is valid when
$(x^3/6)\ll x$ i.e. when $(x^2/6)\ll 1$. Using our maximum expression for the
sine argument in Equation~\ref{eqn:sinmax}, the small-angle
approximation is now valid for:

\begin{align}
&\mathbf{Condition\,\,B} \nonumber \\
&(a/R_{\star})^2 \gg \frac{2}{3} \frac{(1+e)^3}{(1-e)^3}.
\label{eqn:conditionB}
\end{align}

\subsubsection{Summary}

We have now derived the conditions under which the small-angle inverse sine
approximation (Equation~\ref{eqn:conditionA}) and the small-angle sine
approximation (Equation~\ref{eqn:conditionB}) are valid. It is easily shown
that Condition B always leads to a harder constraint on $(a/R_{\star})$, meaning
that Condition A is superfluous. Applying the small-angle approximations to
Equation~\ref{eqn:bobs_ecc} elegantly recovers $b^2$, as 
\citet{investigations:2010} stated. However, the actual limit of this 
approximation is now quantified as:

\begin{align}
\lim_{ (a/R_{\star})^2 \gg [2(1+e)^3]/[3(1-e)^3] } b_{\obs} &= b.
\end{align}

\subsection{Accuracy of the Density Approximation}

With the valid range for assuming $b_{\obs}=b$ now resolved, we may proceed
to finally broach the question as to when Equation~\ref{eqn:photoe1} is valid
i.e. when the analytic model for the photo-eccentric effect can be employed.
The stellar density is trivially computed from $(a/R_{\star})$ and so it is
more pertinent to phrase the question as to what is $(a/R_{\star})_{\obs}$?
\citet{investigations:2010} (Equation~35) showed that (without any 
approximation):

\begin{align}
& \lim_{b_{\obs}\to b} (a/R_{\star})_{\obs}^2 = b^2 + \nonumber\\
\qquad& [(1+p)^2-b^2] \csc^2 \Bigg[ \frac{\varrho_c^2}{\sqrt{1-e^2}} \sin^{-1} \sqrt{ \frac{(1+p)^2-b^2}{ (a/R_{\star}^2 \varrho_c^2 - b^2 } } \Bigg].
\end{align}

At this point \citet{investigations:2010} again invoke an inverse sine and
sine small-angle approximation to simplify the above. However, making these
approximations are equivalent to cases [1] \& [2] of the inverse sine
approximation and cases [1] \& [2] of the sine function approximation made
earlier in this section. Therefore, since we have already assumed Condition B 
(Equation~\ref{eqn:conditionB}) is in effect in order to approximate 
$b_{\obs}=b$, then it necessarily follows that both of these small-angle 
approximations must also be valid. Making these approximations allows for 
significant simplification, yielding the same result as Equation~36 of 
\citet{investigations:2010}:

\begin{align}
& \lim_{ (a/R_{\star})^2 \gg [2(1+e)^3]/[3(1-e)^3] } (a/R_{\star})_{\obs} = (a/R_{\star}) \nonumber\\
\qquad& \sqrt{ \varrho_c^2 \cos^2i + \frac{(1-e^2)\sin^2i}{\varrho_c^2} }.
\end{align}

The final approximation made in \citet{investigations:2010}, which ultimately
yields the photo-eccentric $\Psi$ equation, is that the system is nearly
coplanar. This essentially means that we adopt $\cos i=0$ and $\sin i=1$ in the
above and doing so recovers Equation~\ref{eqn:photoe1}. Explicitly though, the 
assumption may be expressed as:

\begin{align}
\frac{1-e^2}{\varrho_c^2} \sin^2i \gg \cos^2i.
\end{align}

Replacing $\sin^2i$ with $(1-\cos^2i)$ and then replacing $\cos i$ with
$b/[(a/R_{\star})\varrho_c]$ gives:

\begin{align}
(a/R_{\star})^2 \gg \frac{\varrho_c^4+1-e^2}{1-e^2} \frac{b^2}{\varrho_c^2}.
\label{eqn:anotherone}
\end{align}

\begin{figure}
\begin{center}
\includegraphics[width=8.4 cm]{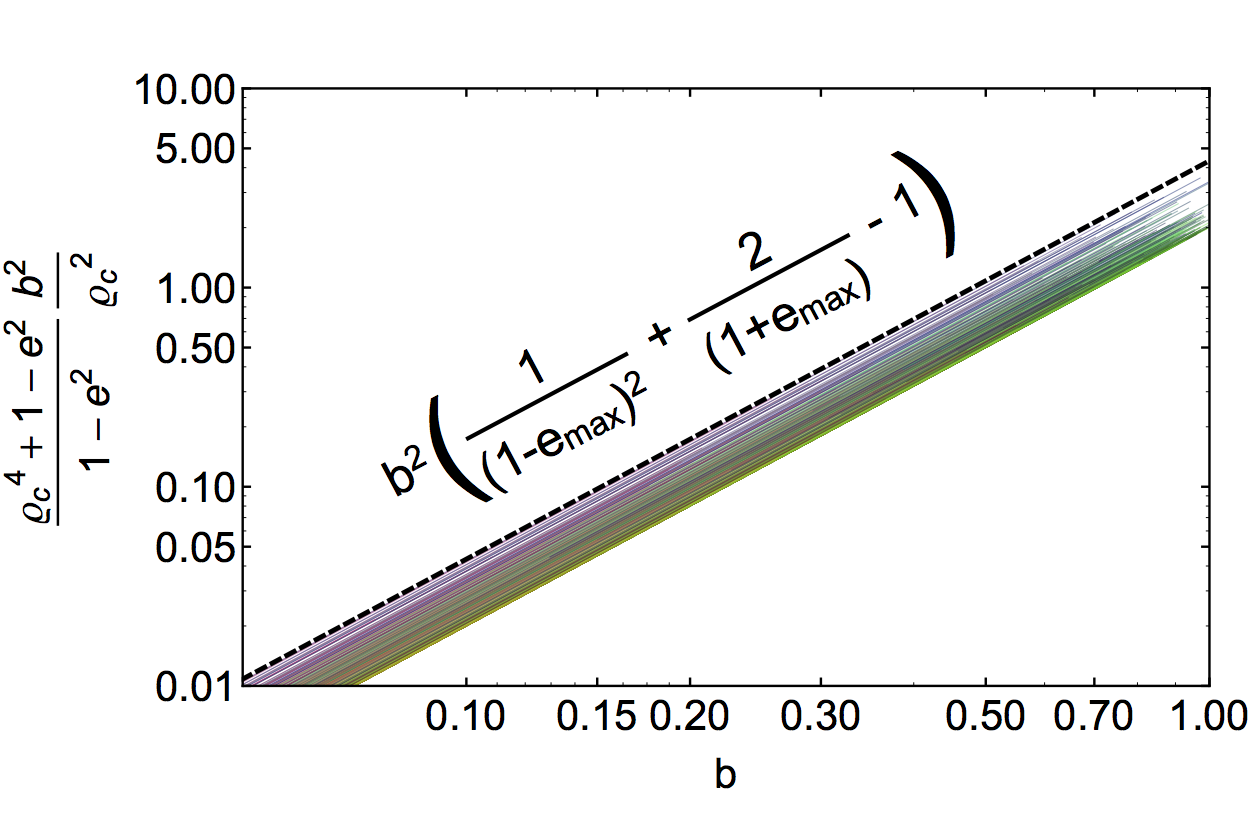}
\caption{\emph{ \textbf{Coplanar approximation investigation:}
Monte Carlo realizations for the constraint on the $(a/R_{\star})^2$ function,
expressed on the $y$-axis and given in Equation~\ref{eqn:anotherone}, with 
respect to $b$. We show 1000 random realizations for the function by drawing
random uniform variates for $p$, $\omega$ and $p$, which respectively define the 
RGB-colouring scheme. The black-dashed lined describes the observed upper limit. 
Simulations produced using $e_{\mathrm{max}}=0.5$, but the upper limits are 
valid for all $0\leq e_{\mathrm{max}}<1$.
}} 
\label{fig:finaltests}
\end{center}
\end{figure}

The function on the RHS depends upon $b$, $e$ and $\omega$ and implicitly $p$
(since $0<b<1-p$). As with earlier, we seek a simple form for the maximum of
the term on RHS by Monte Carlo experiment. In Figure~\ref{fig:finaltests},
we show 1000 random realizations of this function plotted with respect to $b$,
drawing uniform variates for $0\leq p<1$, $0\leq e<e_{\mathrm{max}}$ and 
$0\leq \omega<2\pi$. The exercise reveals that the function is bounded by

\begin{align}
\frac{\varrho_c^4+1-e^2}{1-e^2} \frac{b^2}{\varrho_c^2} \leq b^2 \Big(\frac{1}{(1-e)^2} + \frac{2}{(1+e)} - 1 \Big).
\end{align}

We may now use the above and evaluate it when $b=1$, which maximizes the limit,
to give:

\begin{align}
&\mathbf{Condition\,\,C} \nonumber \\
&(a/R_{\star})^2 \gg \Big(\frac{1}{(1-e)^2} + \frac{2}{(1+e)} - 1 \Big)
\label{eqn:conditionC}
\end{align}

\begin{figure}
\begin{center}
\includegraphics[width=8.4 cm]{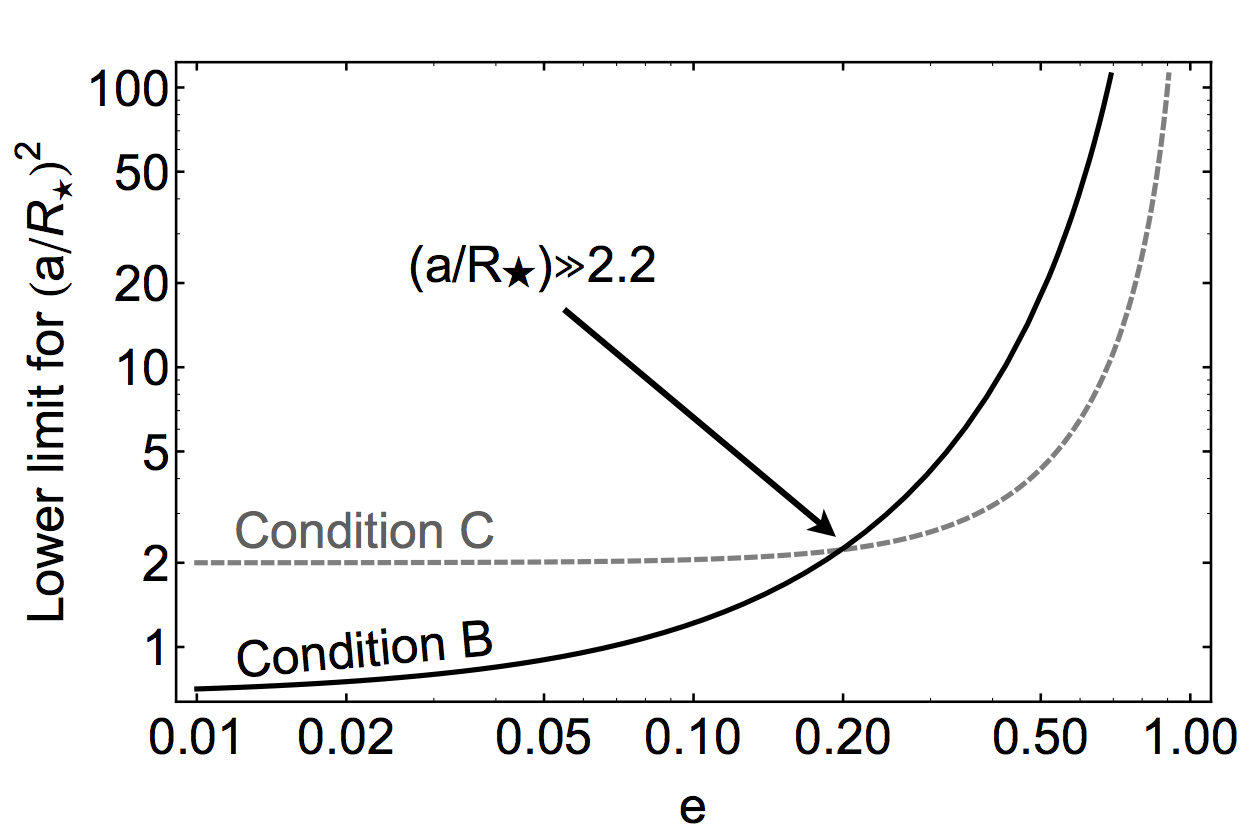}
\caption{\emph{ \textbf{Comparison of conditions B \& C:}
Here we plot the RHS of Equations~\ref{eqn:conditionB}\&\ref{eqn:conditionC}
in order to visualize which of the two conditions dominates. Under the
reasonable assumption that $(a/R_{\star})\gg2$, then Condition B can be seen
to dominate and thus we dub this the analytic photo-eccentric condition.
}} 
\label{fig:conditionsBC}
\end{center}
\end{figure}

Since have assumed Condition B already, it is worth comparing the above to
Equation~\ref{eqn:conditionB}. Plotting the two functions out in 
Figure~\ref{fig:conditionsBC} one sees that unlike the case where we compared
Conditions A \& B, one function does not always dominate over the other.
However, the point of intersection occurs for the constraint that 
$(a/R_{\star})^2\gg2.2$, after which point Condition B dominates. Therefore,
provided we are willing to assume the quite reasonable scenario that 
$(a/R_{\star})^2\gg2$ at all times, then we only need define Condition B as the
defining assumption.

To summarize, the various approximations made in \citet{investigations:2010} may
be explicitly and compactly defined by the following assumption:

\begin{align}
%\mathbf{The\,\,}&\mathbf{Analytic\,\,Photo-eccentric\,\,Condition} \nonumber \\
&(a/R_{\star})^2 \gg \frac{2}{3} \Bigg( \frac{1+e}{1-e} \Bigg)^3,
\end{align}

or equivalently this may be re-expressed in physical dimensions by 
re-writing $(a/R_{\star})$ in terms of the true stellar density

\begin{align}
\Bigg(\frac{P}{\mathrm{days}}\Bigg)^{4/3} \gg 0.101 \Bigg( \frac{\rho_{\star,\tru}}{\mathrm{g\,cm}^{-3}} \Bigg)^{-2/3} \Bigg( \frac{1+e}{1-e} \Bigg)^3.
\label{eqn:conditionX}
\end{align}

Adopting the analytic photo-eccentric condition means one may now re-write 
Equation~\ref{eqn:photoe1} as

\begin{align}
\lim_{ (a/R_{\star})^2 \gg [2(1+e)^3]/[3(1-e)^3] } \Big(\frac{\rho_{\star,\obs}}{\rho_{\star,\tru}}\Big) &= \Psi.
\label{eqn:photoe2}
\end{align}

\section{Derivation of the minimum eccentricity equation}
\label{app:photoeccentric}

In this work, we have presented a new expression for the minimum eccentricity
of an exoplanet (Equation~\ref{eqn:emin}), as a function of the observed and 
true stellar densities ($\rho_{\star,\obs}$ and $\rho_{\star,\tru}$ 
respectively). Here, we present a derivation of this equation. As with the
other derivations in this work, we ignore other effects (e.g. photo-blend,
photo-mass, etc) during the course of this derivation and assume the analytic
photo-eccentric condition (Equation~\ref{eqn:conditionX}) is satisfied. 
Accordingly, the ratio of the observed stellar density to the true stellar 
density follows the expression \citep{investigations:2010}:

\begin{align}
\Big(\frac{\rho_{\star,\obs}}{\rho_{\star,\tru}}\Big) &= \Psi
\label{eqn:investigationseqn}
\end{align}

where

\begin{align}
\Psi &= \frac{(1+e\sin\omega)^{3}}{(1-e^2)^{3/2}}.
\label{eqn:psieqn}
\end{align}

$(\rho_{\star,\obs}/\rho_{\star,\tru})$ can therefore be seen to be a function 
of two parameters, $e$ and $\omega$, meaning that we have one observable and two
unknowns. Progress can be made on this under-constrained problem by considering 
the extrema (i.e. the minima/maxima) of the expression. We will proceed by 
taking the extrema with respect to $\omega$, which is easily achieved by 
computing the derivative with respect to $\omega$.
Solving $\partial(\rho_{\star,\obs}/\rho_{\star,\tru})/\partial\omega=0$ for 
$\omega$ under the condition that $0\leq e<1$ yields two solutions: \
$\omega=\pi/2$ (periapsis transit) and $\omega=3\pi/2$ (apoapsis transit). At 
these extrema, we have

\begin{align}
\lim_{ \omega \to \pi/2 } \Big(\frac{\rho_{\star,\obs}}{\rho_{\star,\tru}}\Big)  &= \Big(-1 + \frac{2}{1+e} \Big)^{-3/2},\\
\lim_{ \omega \to 3\pi/2 } \Big(\frac{\rho_{\star,\obs}}{\rho_{\star,\tru}}\Big) &= \Big(-1 + \frac{2}{1+e}\Big)^{3/2}.
\end{align}

Let us solve the above expressions so that $e$ is the subject:

\begin{align}
\lim_{ \omega \to \pi/2 } e &= \frac{\big(\frac{\rho_{\star,\obs}}{\rho_{\star,\tru}}\big)^{2/3}-1}{\big(\frac{\rho_{\star,\obs}}{\rho_{\star,\tru}}\big)^{2/3}+1},
\label{eqn:e1half}
\end{align}

and

\begin{align}
\lim_{ \omega \to 3\pi/2 } e &= \frac{ (1-\big(\frac{\rho_{\star,\obs}}{\rho_{\star,\tru}}\big)^{2/3})(1-\big(\frac{\rho_{\star,\obs}}{\rho_{\star,\tru}}\big)^{2/3}+\big(\frac{\rho_{\star,\obs}}{\rho_{\star,\tru}}\big)^{4/3}) }{1+\big(\frac{\rho_{\star,\obs}}{\rho_{\star,\tru}}\big)^2}.
\label{eqn:e3half}
\end{align}

If $(\rho_{\star,\obs}/\rho_{\star,\tru})>1$, then Equation~\ref{eqn:e1half} 
yields a positive eccentricity, otherwise it is negative. A negative 
eccentricity of course has no meaning and this can be explained by the fact that 
if $\omega=\pi/2$ then it is impossible to have 
$(\rho_{\star,\obs}/\rho_{\star,\tru})<1$ by simple inspection of 
Equation~\ref{eqn:psieqn}.

The opposite is true for Equation~\ref{eqn:e3half} where if 
$(\rho_{\star,\obs}/\rho_{\star,\tru})<1$ then we arrive at a positive 
eccentricity, otherwise the derived eccentricity is negative. Again, inspection 
of Equation~\ref{eqn:psieqn} reveals that one cannot have a 
$(\rho_{\star,\obs}/\rho_{\star,\tru})>1$ value if $\omega=3\pi/2$.

These two simple observations reveal the applicability of the two expressions.
Specifically, if we have $(\rho_{\star,\obs}/\rho_{\star,\tru})>1$, then we 
should use Equation~\ref{eqn:e1half} and if we have 
$(\rho_{\star,\obs}/\rho_{\star,\tru})<1$ we should use 
Equation~\ref{eqn:e3half}.

Finally, the two extrema can now be interpreted as the minimum eccentricity
of the planet in the two distinct regimes of 
$(\rho_{\star,\obs}/\rho_{\star,\tru})>1$ and 
$(\rho_{\star,\obs}/\rho_{\star,\tru})<1$. This is easily verified by numerical 
tests and the two equations may now be combined into a single term using 
Heaviside Theta functions:

\begin{align}
e_{\min} =& \Big( \lim_{ \omega \to \pi/2 } e \Big) \mathbb{H}\Big[\big(\frac{\rho_{\star,\obs}}{\rho_{\star,\tru}}\big)-1\Big] \nonumber\\
\qquad& + \Big( \lim_{ \omega \to 3\pi/2 } e \Big) \mathbb{H}\Big[1-\big(\frac{\rho_{\star,\obs}}{\rho_{\star,\tru}}\big)\Big],
\end{align}

which we evaluate to be

\begin{align}
e_{\min} =& \Bigg( \frac{ \big(\frac{\rho_{\star,\obs}}{\rho_{\star,\tru}}\big)^{2/3}-1}{\big(\frac{\rho_{\star,\obs}}{\rho_{\star,\tru}}\big)^{2/3}+1} \Bigg) \mathbb{H}\Big[\Big(\frac{\rho_{\star,\obs}}{\rho_{\star,\tru}}\Big)-1\Big] \nonumber \\
\quad& + \Bigg( \frac{ (1-\big(\frac{\rho_{\star,\obs}}{\rho_{\star,\tru}}\big)^{2/3})(1-\big(\frac{\rho_{\star,\obs}}{\rho_{\star,\tru}}\big)^{2/3}+\big(\frac{\rho_{\star,\obs}}{\rho_{\star,\tru}}\big)^{4/3}) }{1+\big(\frac{\rho_{\star,\obs}}{\rho_{\star,\tru}}\big)^2} \Bigg) \nonumber\\
\qquad& \mathbb{H}\Big[1-\Big(\frac{\rho_{\star,\obs}}{\rho_{\star,\tru}}\Big)\Big],
\end{align}

Note that $e_{\min}$ is purely a function of 
$(\rho_{\star,\obs}/\rho_{\star,\tru})$ and no other terms. It is
therefore possible to analytically calculate the uncertainty on $e_{\min}$
using quadrature:

\begin{align}
\sigma_{e_{\min}} &= \frac{4}{3} \Big(\frac{\rho_{\star,\obs}}{\rho_{\star,\tru}}\Big)^{-1/3}\Bigg(1+\Big(\frac{\rho_{\star,\obs}}{\rho_{\star,\tru}}\Big)^{2/3}\Bigg)^{-2} \sigma_{ \rho_{\star,\obs}/\rho_{\star,\tru} },
\end{align}

where $\sigma_{e_{\min}}$ and $\sigma_{ \rho_{\star,\obs}/\rho_{\star,\tru} }$ 
are the uncertainties on the minimum eccentricity and 
$(\rho_{\star,\obs}/\rho_{\star,\tru})$ observable respectively.

\section{Derivation of the Photo-timing Effect}
\label{app:phototiming}

Consider $N\gg1$ transits exhibiting periodic transit timing variations (TTV)
with a period much less than the baseline of observations, such that a large
number of TTV oscillations have occurred over the span of the time series. If
one was unaware of these TTVs, the default assumption would be to fit a linear 
ephemeris model through the transits. This is equivalent to folding the
transits upon a particular linear ephemeris. Figure~\ref{fig:TTVplot} 
illustrates 100 transits exhibiting sinusoidal TTVs folded in this way. The 
displacement of each transit from the central folded time means that when we 
average the light curves to produce a composite signal, this composite signal 
displays a different morphology to the individual transits. Specifically, the 
first and fourth contact points are pulled outwards and the second and third 
contact points are pulled inwards. This has the effect of mimicking a more 
grazing event and thus increases $b$. Since $b$ is inversely correlated to 
$(a/R_{\star})$ and thus $\rho_{\star}$, one should anticipate that unaccounted 
for TTVs will produce an artificially lower $\rho_{\star}$ value. This may be 
formally proved here by considering the effect on the contact points and 
following the method outlined in \citet{binning:2010}. For a peak-to-peak TTV 
amplitude of $2A_{\mathrm{TTV}}$, the contact points of the composite signal 
appear shifted by:

\begin{align}
t_{I,\obs} &= t_{I,\tru} - A_{\mathrm{TTV}},\\
t_{II,\obs} &= t_{II,\tru} + A_{\mathrm{TTV}},\\
t_{III,\obs} &= t_{III,\tru} - A_{\mathrm{TTV}},\\
t_{IV,\obs} &= t_{IV,\tru} + A_{\mathrm{TTV}}.\\
\end{align}

Together, these change the apparent transit durations, $T_{23}$ and $T_{14}$,
to:

\begin{align}
T_{23,\obs} &= T_{23,\tru} - 2A_{\mathrm{TTV}},\\
T_{14,\obs} &= T_{14,\tru} + 2A_{\mathrm{TTV}}.
\label{eqn:phototimingdurations}
\end{align}

Extreme scenarios can cause $T_{23,\obs}<0$ thus mimicking a grazing event, 
which we do not consider here. In practice, such large TTVs are easily detected
and thus unlikely to go unaccounted for. Therefore, we may consider the transit
depth to be unaffected and thus $p_{\obs}=p_{\tru}$. 

In order to compute the deviation in $(\rho_{\star,\obs}/\rho_{\star,\tru})$
from unity due the photo-timing effect, one may follow the methodology outlined
in \S\ref{sub:method} and feed Equation~\ref{eqn:phototimingdurations} into
Equations~\ref{eqn:bseager}, \ref{eqn:aRseager} and \ref{eqn:rhoseager}.
Performing these steps yields highly elaborate expressions for 
$(\rho_{\star,\obs}/\rho_{\star,\tru})$ and rather than formally stating the 
full equation (requiring many lines), meaningful insights may be drawn by 
plotting the resulting function for various impact parameters. 

\begin{figure}
\begin{center}
\includegraphics[width=8.4 cm]{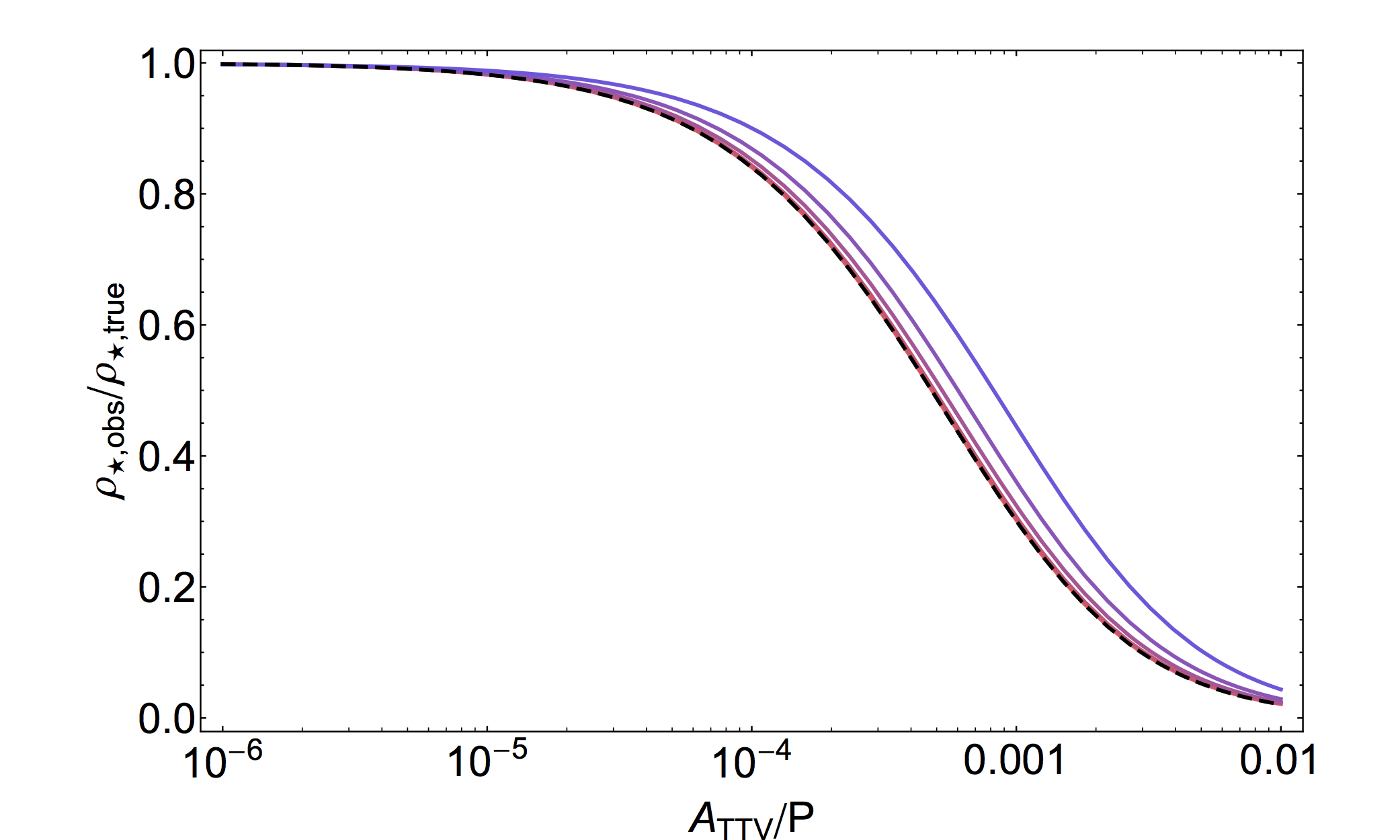}
\caption{\emph{The effect of unaccounted for transit timing variations (x-axis) 
on the observed mean stellar density (y-axis) from a composite transit light 
curve. From red to blue we show iso-$b$ contours of $b=0.0, 0.2, 0.4, 0.6$ \&
$0.8$ respectively. The black-dashed line shows the result of our approximate
expression in the $b=0$ limit (Equation~\ref{eqn:TTVapprox}). Realizations
computed using $P=10$\,days, $\rho_{\star} = \rho_{\odot}$ and $p=0.1$.}} 
\label{fig:TTV_vs_b}
\end{center}
\end{figure}

In Figure~\ref{fig:TTV_vs_b}, we plot the ratio 
$(\rho_{\star,\obs}/\rho_{\star,\tru})$ as a function of $(A_{\mathrm{TTV}}/P)$ 
for several iso-$b$ contours. The plot reveals that the maximal error in 
$(\rho_{\star,\obs}/\rho_{\star,\tru})$ occurs for $b=0$ and so we may continue 
by focussing our efforts on this case and interpreting it as the maximal 
deviation. We find that for $(A_{\mathrm{TTV}}/P)\simeq10^{-5}$ the 
$(\rho_{\star,\obs}/\rho_{\star,\tru})$ term is deviant by 1\%.

\begin{align}
\lim_{ b \to 0 } \Big( \frac{ \rho_{\star,\obs} }{ \rho_{\star,\tru} }\Big) &=
\Big(\frac{3\pi}{G P^2}\Big) \nonumber \\
\qquad& \times \csc^3\Big[n A_{\mathrm{TTV}} + (1+p) (a/R_{\star})_{\tru}^{-1} \Big] \Bigg[ (1+p)^2 \nonumber \\
\qquad& + \cos[n A_{\mathrm{TTV}} + (1+p) (a/R_{\star})_{\tru}^{-1}] \nonumber\\
\qquad& \times \Bigg( (1+p)^2 \frac{ \sin^2[n A_{\mathrm{TTV}} - (1-p) (a/R_{\star})_{\tru}^{-1}] }{ \sin^2[n A_{\mathrm{TTV}} + (1+p) (a/R_{\star})_{\tru}^{-1}] } \nonumber \\
\qquad& - (1-p)^2 \Bigg) \Bigg( 1 - \nonumber \\
\qquad& \frac{ \sin^2[n A_{\mathrm{TTV}} - (1-p) (a/R_{\star})_{\tru}^{-1}] }{ \sin^2[n A_{\mathrm{TTV}} + (1+p) (a/R_{\star})_{\tru}^{-1}] } \Bigg)^{-1} \Bigg]^{3/2},
\label{eqn:TTVexact}
\end{align}

where $n=2\pi/P$. Making small-angle approximations of the various trigonometric
terms, allows for considerable simplification of this equation:

\begin{align}
\lim_{ b \to 0 } \Big( \frac{ \rho_{\star,\obs} }{ \rho_{\star,\tru} }\Big) &\simeq \Bigg( \frac{p}{p+n A_{\mathrm{TTV}} (a/R_{\star})_{\tru}} \Bigg)^{3/2}
\label{eqn:TTVapprox}
\end{align}

Similar small-angle approximations have been made previously in this work 
in Appendix~\ref{app:psi}, where we derived the exact parameter range of the
approximation's validity. In the case of the photo-timing effect derivation
presented here, we need only concern ourselves with the $e\rightarrow0$
limit, since our derivations consider each AP effect in isolation. The only
remaining difference now is that we require $2A_{\mathrm{TTV}} \ll T_{23}$
for the exact same approximation to be valid. Let us invoke this reasonable
assumption since any large TTVs which break this condition should be easily
detected and compensated for and the photo-timing effect concerns itself
with clandestine timing variations. In Appendix~\ref{app:psi}, we found that two 
conditions were required for the small-angle approximations; conditions B \& C, 
given by Equations~\ref{eqn:conditionB} \& \ref{eqn:conditionC}. Using those 
same expressions, but setting $e=0$ as appropriate for the photo-timing 
derivation being considered here, we find that such an approximation is
generally valid if:

\begin{align}
%&\mathbf{The\,\,Analytic\,\,Phototiming\,\,Condition} \nonumber \\
&(a/R_{\star})^2 \gg 2,\\
&2A_{\mathrm{TTV}} \ll T_{23}.
\end{align}

which can be considered to be true for the vast majority of orbital 
configurations. As visible from Figure~\ref{fig:TTV_vs_b}, the approximation 
given by Equation~\ref{eqn:TTVapprox} does an excellent
job of reproducing the behavior of the exact solution for $b=0$.  This equation
is also highly practical in estimating the error in 
$(\rho_{\star,\obs}/\rho_{\star,\tru})$ when some upper limit on the TTVs has
been derived, since the $b=0$ limit is the most conservative case and in
general the derived $b$ will over-estimated and thus unreliable anyway due
to the TTV smearing. This TTV smearing imposes a fundamental limit on the 
precision at which one can measure $(\rho_{\star,\obs}/\rho_{\star,\tru})$.

To make Equation~\ref{eqn:TTVapprox} of even greater practical value to
observers, it is useful to replace $(a/R_{\star})_{\tru}$ with 
$\rho_{\star,\tru}$ since this parameter is more directly inferred from an
independent measure of the star. Further, in the case of no detected TTVs,
the $A_{\mathrm{TTV}}$ may be replaced with the upper limit on the TTV
amplitude and the LHS may be interpretted as the uncertainty in 
$(\rho_{\star,\obs}/\rho_{\star,\tru})$:

\begin{align}
\sigma_{(\rho_{\star,\obs}/\rho_{\star,\tru})}^{\mathrm{max}} &= 1 - \Bigg( 1+ \frac{2\pi^{2/3}}{3^{1/3}} \frac{G^{1/3}\rho_{\star,\tru}^{1/3}}{p P^{1/3}} \sigma_{A_{\mathrm{TTV}}} \Bigg)^{-3/2},
\label{eqn:TTVerror1}
\end{align}

where

\begin{align}
\sigma_{(\rho_{\star,\obs}/\rho_{\star,\tru})} \leq \sigma_{(\rho_{\star,\obs}/\rho_{\star,\tru})}^{\mathrm{max}}
\end{align}

and $\sigma_{A_{\mathrm{TTV}}}$ is the 1\,$\sigma$ upper limit on the presence
of TTVs. For $N\gg1$ transits observed with an approximately constant timing
precision of $\sigma_{\tau}$, one expects the standard deviation of the
TTV points in the absence of a signal to be $\sigma_{\tau}$. The uncertainty on
this prediction (i.e. the standard deviation of the standard deviation) is
given by $\sigma_{\tau}/\sqrt{2(N-1)}$ assuming normally distributed errors.
The 1\,$\sigma$ maximum standard deviation can then be compared to that expected
from an embedded sinusoid within the data which could cause a standard deviation
of $\sqrt{\sigma_{\tau}^2+(A^2/2)}$:

\begin{align}
\sqrt{\sigma_{\tau}^2+(A^2/2)} &= \sigma_{\tau} + \sigma_{\tau}/\sqrt{2(N-1)}.
\end{align}

Solving for $A$ gives $\sigma_{A_{\mathrm{TTV}}}$ as

\begin{align}
\sigma_{A_{\mathrm{TTV}}} &= \sigma_{\tau} \sqrt{ \frac{1}{N-1} + \sqrt{\frac{8}{N-1}} },\\
\lim_{N\gg1} \sigma_{A_{\mathrm{TTV}}} &\simeq \sigma_{\tau} \Big( \frac{8}{N} \Big)^{1/4}.
\end{align}

We may now plug the above result into Equation~\ref{eqn:TTVerror1}. Further,
we take an approximate estimate of the $\rho_{\star,\tru}$ term on the RHS of
Equation~\ref{eqn:TTVerror1} to be equal to $\rho_{\star,\obs}$. Finally,
we assume that the fractional error is much less than unity to simplify the
expression to

\begin{align}
\sigma_{(\rho_{\star,\obs}/\rho_{\star,\tru})}^{\mathrm{max}} &\simeq 7.5 \frac{G^{1/3}\rho_{\star,\obs}^{1/3}}{p P^{1/3}} \frac{\sigma_{\tau}}{N^{1/4}}.
\label{eqn:TTVerror2}
\end{align}

In a Taylor expansion of $[1-(1+x)^{3/2}]$, we require $x\ll0.8$ for 
Equation~\ref{eqn:TTVerror2} to be applicable, which corresponds to:

\begin{align}
\Bigg(\frac{P}{\mathrm{days}}\Bigg)^{1/3} \gg \frac{1}{150 p} \Bigg(\frac{A_{\mathrm{TTV}}}{\mathrm{seconds}}\Bigg) \Bigg(\frac{\rho_{\star,\tru}}{\mathrm{g\,cm}^{-3}}\Bigg)^{-1/3}.
\end{align}

\section{Derivation of the Photo-duration Effect}
\label{app:photoduration}

Consider a planet undergoing periodic, low-amplitude velocity-induced transit 
duration variations (TDV-Vs). By periodically increasing/decreasing the velocity 
of a planet, one expects the transit duration to scale inversely. In 
Figure~\ref{fig:TDVplot}, the effect is illustrated on the composite light 
curve.

The outcome of unaccounted TDV-Vs is similar to that of unaccounted TTVs. 
Namely, the first and fourth contacts are pulled outwards and the second and 
third contacts are pulled inwards. Thus one should expect unaccounted TDV-Vs to 
cause one to underestimate the stellar density, like TTVs. 

In what follows we consider the effect of a sinusoidal velocity variation via

\begin{align}
v(t) &= v_0 [ 1 - A_{\mathrm{TDV}} \sin (2\pi t/P_{\mathrm{TDV}}) ].
\end{align}

However, it is important to note the derivation is general for any periodic
waveform and in this sense our models defines $A_{\mathrm{TDV}}$ as half of
the peak-to-peak velocity variation amplitude. Since $T_{ _{23}^{14} }$ is 
linearly inversely proportional to the velocity, $v$, then we have:

\begin{align}
T_{ _{23}^{14} }(t) = T_{ _{23}^{14},0 } [ 1 - A_{\mathrm{TDV}} \sin (2\pi t/P_{\mathrm{TDV}}) ]^{-1}.
\end{align}

If $A_{\mathrm{TDV}}\ll1$, then we have:

\begin{align}
T_{ _{23}^{14} }(t) \simeq T_{ _{23}^{14},0 } [ 1 + A_{\mathrm{TDV}} \sin (2\pi t/P_{\mathrm{TDV}}) ]
\end{align}

In such a case, one can show that the composite contact points are shifted by

\begin{align}
t_{I,\obs} &= t_{I,\tru} - A_{\mathrm{TDV}} T_{14,0},\\
t_{II,\obs} &= t_{II,\tru} + A_{\mathrm{TDV}} T_{23,0},\\
t_{III,\obs} &= t_{III,\tru} - A_{\mathrm{TDV}} T_{23,0},\\
t_{IV,\obs} &= t_{IV,\tru} + A_{\mathrm{TDV}} T_{14,0}.\\
\end{align}

Together, these change the observed transit durations, $T_{23}$ and $T_{14}$,
to:

\begin{align}
T_{23,\obs} &= T_{23,\tru} - 2 A_{\mathrm{TDV}} T_{23,\tru},\\
T_{14,\obs} &= T_{14,\tru} + 2 A_{\mathrm{TDV}} T_{14,\tru}.
\end{align}

One may now proceed to derive the effect on $\rho_{\star,\obs}$ as we did before
for the photo-timing effect. However, unlike the photo-timing effect, we find
that a simple form of the equation is possible for all $b$ values, given by:

\begin{align}
&\Big(\frac{\rho_{\star,\obs}}{\rho_{\star,\tru}}\Big) = \nonumber\\
&\Bigg( \frac{ (a/R_{\star})^2 p + 4 A_{\mathrm{TDV}}^2 b^2 p + 2 A_{\mathrm{TDV}} [ (1-p^2)^2-b^2(1+p^2) ] }{ (a/R_{\star})^2 [p+4A_{\mathrm{TDV}}^2p+2A_{\mathrm{TDV}} (1+p^2-b^2)] } \Bigg)^{3/2},
\label{eqn:photodurationapprox}
\end{align}

where $(a/R_{\star})$ is $(a/R_{\star})_{\tru}$ and can be estimated as 
$[(G P^2 \rho_{\star})/(3\pi)]^{1/3}$. 
As with the previous derivations, the above required making similar small-angle 
approximations to those made in Appendix~\ref{app:psi}. These approximations are
valid here too under the already made assumption that $A_{\mathrm{TDV}}\ll1$,
meaning we assume:

\begin{align}
%&\mathbf{The\,\,Analytic\,\,Photoduration\,\,Condition} \nonumber \\
&(a/R_{\star})^2 \gg 2,\\
&A_{\mathrm{TDV}} \ll 1.
\end{align}

\bsp

\label{lastpage}

\end{document}